\documentclass[12pt]{article}
\usepackage{lineno}
\usepackage{amsthm}
\usepackage{here}
\usepackage{natbib}
\usepackage{lscape}
\usepackage{longtable}
\usepackage{bm,graphicx,graphics,url,psfrag}
\usepackage{lipsum} 
\usepackage{amsthm,amsmath,amssymb,amsfonts,mathrsfs,mathtools}
\usepackage{breakcites}
\usepackage{subfigure}
\usepackage{psfrag}

\usepackage[paperwidth=26cm]{geometry}

\usepackage{multirow}

\usepackage{listings}
\usepackage{color}

\newtheorem{Theorem}{Theorem}[section]

\newtheorem{Lemma}{Lemma}[section]
\newtheorem{Proposition}{Proposition}[section]
\newtheorem{Remark}{Remark}[section]

\definecolor{codegreen}{rgb}{0,0.6,0}
\definecolor{codegray}{rgb}{0.5,0.5,0.5}
\definecolor{codepurple}{rgb}{0.58,0,0.82}
\definecolor{backcolour}{rgb}{0.95,0.95,0.92}

\lstdefinestyle{mystyle}{
    backgroundcolor=\color{backcolour},
    commentstyle=\color{codegreen},
    keywordstyle=\color{magenta},
    numberstyle=\tiny\color{codegray},
    stringstyle=\color{codepurple},
    basicstyle=\footnotesize,
    breakatwhitespace=false,
    breaklines=true,
    captionpos=b,
    keepspaces=true,
    numbers=left,
    numbersep=5pt,
    showspaces=false,
    showstringspaces=false,
    showtabs=true,
    tabsize=2
}

\title{\bf On a bivariate Birnbaum-Saunders distribution parameterized by its means: features, reliability analysis and application}
\author{
\textbf{\normalsize \text{Helton Saulo}$^{1}$,
\text{Jeremias Le\~ao}$^{2}$,
\text{Roberto Vila}$^{1}$,
\text{Victor Leiva}$^{3}$\footnote{Corresponding author: Victor Leiva. Email: victorleivasanchez@gmail.com.}\,\,,
\text{Vera Tomazella}$^{4}$}
\\[0.20cm]
{\small $^{1}$Department of Statistics, Universidade de Bras\'{i}lia, Brazil}\\[-0.1cm]
{\small $^{2}$Department of Statistics, Universidade Federal do Amazonas, Brazil}\\[-0.1cm]
{\small $^{3}$School of Industrial Engineering, Pontificia Universidad Cat\'{o}lica de Valpara\'{i}so, Chile}\\[-0.1cm]
{\small $^{4}$Department of Statistics, Universidade Federal de S\~ao Carlos, Brazil}\\[-0.1cm]
}

\begin{document}
\maketitle

\begin{abstract}
Birnbaum-Saunders models have been widely used to model
positively skewed data. In this paper, we introduce a bivariate
Birnbaum-Saunders distribution which has the means as parameters. We
present some properties of the univariate and bivariate Birnbaum-Saunders models.
We discuss the maximum likelihood and modified moment estimation of the model
parameters and associated inference. A simulation study is conducted to evaluate
the performance of the maximum likelihood and modified moment estimators. The probability
coverages of confidence intervals are also discussed. Finally, a real-world data analysis
is carried out for illustrating the proposed model.

\end{abstract}

\section{Introduction}

The Birnbaum-Saunders (BS) distribution was proposed by \cite{bs:69a} motivated by problems
of vibration in commercial aircrafts that caused fatigue in materials. Although, in principle,
its origin is for modeling equipment lifetimes subjected to dynamic loads, the BS distribution
has been widely studied and applied in many applied fields including, for example, engineering,
business, economics, medicine, atmospheric contaminants, finance and quality control; see \cite{jk:03},
\cite{bll:07}, \citet{b:10}, \cite{vpl:11}, \cite{plbl:12}, \cite{slzm:13}, \cite{mbls:13}, \cite{lmsr:14},
\cite{leaolst:17,leaolst:16}, and references therein. The interested reader on the BS distribution
is refereed to \cite{jkb:95,l:15}. These works present a full review about this model. The BS distribution
has been used quite effectively to model positively skewed data, especially lifetime data and crack growth data. This
distribution is related to the normal distribution and has interesting properties.

Some works on reparameterized versions of the BS distribution were proposed by \cite{ablv:08},
\cite{ltw:10} and \cite{scla:12,sclb:16}. In particular, the work of \cite{scla:12} proposed
several parameterizations of the BS distribution, which allow diverse features
of data modeling to be considered.  One of such parameterizations is indexed by the parameters $\mu$
and $\delta$, where $\mu>0$ is a scale parameter and the mean of the distribution, whereas $\delta>0$
is a shape and precision parameter. The notation $T\sim \textrm{RBS}(\mu,\delta)$ is used when an random
variable (RV) follows a reparameterized BS (RBS) distribution; more details about this model is found in
Section \ref{sec:2}.

The bivariate version of the BS distribution was proposed by \cite{kbj:10}, where the authors
discussed maximum likelihood (ML) estimation and modified moment (MM) estimation of the model parameters.
Recently, \cite{kkj:15} observed that the bivariate BS model proposed by \cite{kbj:10} can be
written as the weighted mixture of bivariate inverse Gaussian distribution \citep{kocherlakota:86}
and its reciprocals. They also introduced a mixture of two bivariate BS distributions and discussed
its various properties. \cite{kbj:13} extended to the multivariate case the generalized BS distribution
introduced by \cite{dl:05}. Other bivariate and multivariate distributions related to the BS model
can be found in \cite{vbz:14a,vbz:14b}, \cite{k:15a,k:15b} and \cite{jk:15}.

In this context, the primary objective of this paper is to introduce a bivariate RBS (BRBS)
distribution based on the RBS distribution proposed by \cite{scla:12}.
The secondary objectives are: (i) to present the reliability measure for the BRBS
model and obtain the monotonicity of the hazard rate (HR); (ii) to discuss some unimodality properties of
the BRBS model; (iii) to derive the ML estimators and MM estimators of the unknown parameters
as well as their asymptotic properties; (iv) to evaluate the performance of
the ML and MM estimators via Monte Carlo (MC) simulations; and (v) to illustrate the potential
applications by using real data.

The rest of the paper proceeds as follows. In Section~\ref{sec:2},
we describe briefly the BS and RBS distributions and their related properties. In Section~\ref{sec:3},
we introduce the BRBS distribution and discuss some of its properties. In Section~\ref{sec:4}, we present the ML
and MM estimators of the unknown parameters and their corresponding asymptotic results. In Section~\ref{sec:5}, an evaluation of the ML and MM estimators using MC simulation
is shown, as well as an illustrative example by using a real data set. Finally, in Section~\ref{sec:6}, we provide some concluding remarks
and also point out some problems worthy of further study.

\section{BS distributions}\label{sec:2}

\subsection{The BS distribution}

The BS distribution is related to the normal distribution by means of the stochastic representation
\begin{align*}
T=\frac{\beta}{4}\left[\alpha{Z} + \sqrt{(\alpha{Z})^2+4}\right]^{2},
\end{align*}
where $\alpha>0$ and $\beta>0$ are shape and scale parameters,
respectively, $Z$ is a RV following a standard normal
distribution $Z \sim \textrm{N}(0, 1)$, such that $T$ is BS distributed
with notation $T\sim\textrm{BS}(\alpha, \beta)$. The probability density
function (PDF) of $T$ is given by
\begin{equation}\label{sec1:01}
f(t;\alpha,\beta)
= \frac{1}{\sqrt{2 \pi}} \,
\exp \left ( -\frac{1}{2 \alpha^2} \left [ \frac{t}{\beta} + \frac{\beta}{t} -2\right ]\right)\,
\frac{t^{-3/2} (t + \beta)}{2 \alpha \sqrt{\beta}}, \quad t > 0.
\end{equation}

The mean and variance of $T$ are given
by $\textrm{E}[T]=\beta[1+\alpha^2/2]$ and $\textrm{Var}[T]=(\alpha\beta)^2(1+5\alpha^2/4)$,
respectively. The scale parameter $\beta$
is also the median of the distribution. The BS distribution
holds the reciprocal property,
that is, $1/T$ has the same distribution of $T$ with the
parameter $\beta$ replaced by $1/\beta$, $1/T\sim\textrm{BS}(\alpha, 1/\beta)$, which
implies
\begin{equation*}\label{sec1:02}
\textrm{E}\left[\frac{1}{T}\right]
=
\beta\left(1+\frac{\alpha^2}{2}\right),
\quad
\textrm{ Var}\left[\frac{1}{T}\right]
=
\frac{\alpha^2}{\beta^2}\left(1+\frac{5}{4}\alpha^2\right).
\end{equation*}

\subsection{The RBS distribution}\label{pdf:RBS}

\begin{figure}[htbp]
\centering
\psfrag{fu}[c][c]{\scriptsize{$f(t; \mu, \delta)$}}
\psfrag{su}[c][c]{\scriptsize{$S(t;\mu, \delta)$}}
\psfrag{hu}[c][c]{\scriptsize{$h(t; \mu, \delta)$}}
\psfrag{u}[c][c]{\scriptsize{$u$}}
\psfrag{0}[c][c]{\scriptsize{0}}
\psfrag{1}[c][c]{\scriptsize{1}}
\psfrag{2}[c][c]{\scriptsize{2}}
\psfrag{3}[c][c]{\scriptsize{3}}
\psfrag{4}[c][c]{\scriptsize{4}}
\psfrag{5}[c][c]{\scriptsize{5}}
\psfrag{0.0}[c][c]{\scriptsize{0.0}}
\psfrag{0.2}[c][c]{\scriptsize{0.2}}
\psfrag{0.4}[c][c]{\scriptsize{0.4}}
\psfrag{0.5}[c][c]{\scriptsize{0.5}}
\psfrag{0.6}[c][c]{\scriptsize{0.6}}
\psfrag{0.8}[c][c]{\scriptsize{0.8}}
\psfrag{1.0}[c][c]{\scriptsize{1.0}}
\psfrag{1.5}[c][c]{\scriptsize{1.5}}
\psfrag{2.0}[c][c]{\scriptsize{2.0}}
\psfrag{2.5}[c][c]{\scriptsize{2.5}}
\psfrag{a}[l][c]{\scriptsize{$\delta=0.01$}}
\psfrag{b}[l][c]{\scriptsize{$\delta=0.10$}}
\psfrag{c}[l][c]{\scriptsize{$\delta=1.00$}}
\psfrag{d}[l][c]{\scriptsize{$\delta=2.00$}}
\psfrag{e}[l][c]{\scriptsize{$\delta=5.00$}}
{\includegraphics[height=5.5cm,width=5.5cm,angle=-90]{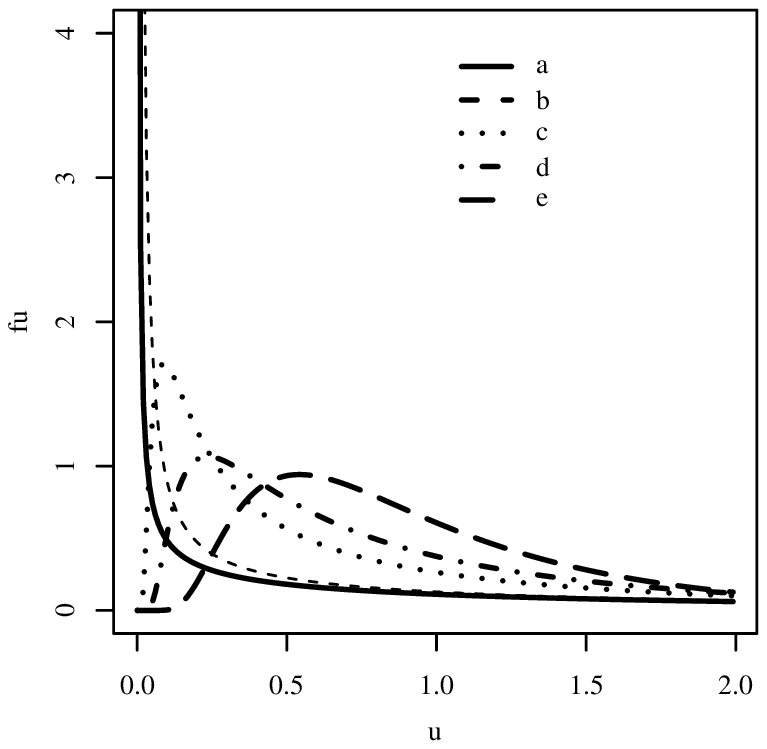}}
{\includegraphics[height=5.5cm,width=5.5cm,angle=-90]{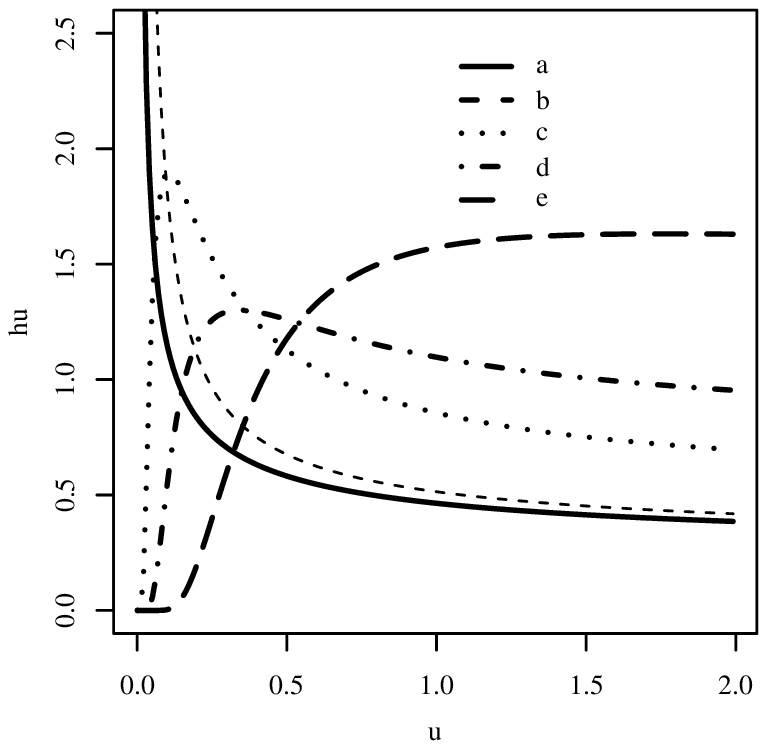}}
\caption{\small PDF (left) and HR (right) plots of the RBS
model for $\mu = 1$ and indicated values of $\delta$.}
\label{fig:rbs}
\end{figure}

The RBS distribution is indexed by the parameters
$\mu=\beta(1+\alpha^{2}/2)$ and $\delta=2/\alpha^{2}$, where $\alpha$
and $\beta$ are the original BS parameters of \eqref{sec1:01}, and $\mu,\delta>0$
are the scale and shape (precision) parameters, respectively. If $T\sim \textrm{RBS}(\mu,\delta)$,
then its PDF, for $t>0$, is given by
\begin{align}\label{eq01}
f(t;\mu, \delta)=
\frac{\exp(\delta/2)\sqrt{\delta+1}}{4\,t^{3/2}\,\sqrt{\pi\mu}}\left(t+\frac{\delta
\mu}{\delta+1}\right)
\exp\left(-\dfrac{\delta}{4}\left[\dfrac{t(\delta+1)}{\delta\mu}+
\dfrac{\delta\mu}{t(\delta+1)}\right]\right),
\end{align}
and its cumulative distribution function (\textrm{CDF}) is denoted by
$F(t;\mu, \delta)$.
From \eqref{eq01}, the survival function (\textrm{SF})
and hazard rate (\textrm{HR}) function are given by
\begin{align*}
S(t;\mu,\delta)
&=
\dfrac{1}{2}
\Phi\left(\displaystyle \dfrac{t+\delta(t-\mu)}{ 2\sqrt{t(1+\delta)\mu}}\right),
\\
h(t;\mu, \delta)
&=
\frac{\exp\left(\displaystyle -\frac{(-\delta  \mu +\delta  t+t)^2}{4 (\delta +1) \mu  t}\right)
(\delta  \mu +\delta  t+t)}{2 \sqrt{\pi \mu (\delta +1)} \sqrt{\mu } \,t^{3/2} \, \Phi\left(\dfrac{ t+\delta(t-\mu)}{2\sqrt{t(1+\delta)\mu}}\right)},
\end{align*}
respectively,
where $\Phi(\cdot)$ is the \textrm{CDF} of the standard normal distribution.
Figure~\ref{fig:rbs} displays some shapes for the \textrm{PDF} and HR of $T\sim \textrm{RBS}(1,\delta)$.

Considering the function
\begin{align}\label{skewe-function}
a(t;\alpha,\beta)={1\over \alpha}
\left[\sqrt{t\over \beta}-\sqrt{\beta\over t}\right],
\quad \alpha,\beta >0,
\end{align}
and denoting $a(t)=a(t;\alpha,\beta)$, where
$\alpha=\sqrt{2/\delta}$ and $\beta=\mu\delta/(\delta+1)$,
expressions for the first, second and third derivatives of $a(\cdot)$ are given by
\begin{align}\label{rem-us}
a'(t)=
{1\over 2\alpha}&\left[{1\over \sqrt{\beta t}}+{1\over t}{\sqrt{\beta\over t}}\right],
\quad
a''(t)=-{1\over 4\alpha t}\left[{1\over \sqrt{\beta t}}+{3\over t}{\sqrt{\beta\over t}}\right],
\quad
a'''(t)={3\over 8\alpha t^2}\left[{1\over \sqrt{\beta t}}+{5\over t}{\sqrt{\beta\over t}}\right].
\end{align}
Note that
$
f(t;\mu, \delta)
=
\phi\left(a(t)\right)
a'(t)
$
where $\phi(\cdot)$ denotes the \textrm{PDF} of the standard normal distribution.
The function $a(\cdot)$ is a bijection of $\mathbb{R}^+$ to $\mathbb{R}$ and has inverse,
denoted by $a^{-\perp}(\cdot)$, given by
\[
a^{-\perp}(s)=\frac{\beta}{4}\left[\alpha{s} + \sqrt{(\alpha{s})^2+4}\right]^{2},
\quad s\in\mathbb{R}.
\]

Some properties of the RBS distribution are the following. Let
$\alpha=\sqrt{2/\delta}$ and $\beta=\mu\delta/(\delta+1)$.
The RBS distribution satisfies the following properties:
\begin{enumerate}
\item The PDF of the RBS distribution has at most one mode, see Proposition 2.7 in \cite{vlsns:17}.
\item If $T\sim \textrm{RBS}(\mu,\delta)$, the moments about the origin of $T$
are given by
	\subitem
	$
	\textrm{E}[T]
	=
	{\beta\over 2}
	\left(\alpha^2+2\right)
	=
	\mu,
	$
	\subitem
	$
	\textrm{E}[T^2]
	=
	{\beta^2\over 2}
	\left(4\alpha^2+3\alpha^4+2\right)
	=
	\mu^2\left(1+{\delta+5\over (\delta+1)^2}\right),
	$
	\quad and in general
	\subitem
	$
	\textrm{E}[T^n]
	=
	\left({\beta\over 2}\right)^n
	\sum_{k=0}^{n}\binom{n}{k}
	\textrm{E}
	\left[(\alpha^2Z^2+2)^k(\alpha Z\sqrt{(\alpha Z)^2+4})^{n-k} \right],
	$

where $Z\sim \textrm{N}(0,1)$.
Then,
	\subitem
	$
	\textrm{Var}[T]
	=
	\mu^2\, {\delta+5\over (\delta+1)^2}.
	$

Since $\textrm{E}\left[|Z|^n\right]<+\infty$, the $n$th moment
$\textrm{E}[T^n]$ always exists. As mentioned, $\delta$ can be interpreted as a precision parameter, that is,
for fixed values of $\mu$, when $\delta \to \infty$, the variance of $T$
tends to zero. Also, for fixed $\mu$, if $\delta \to 0$, then $\textrm{Var}[T] \to 5\mu^2$.

\item If $T\sim \textrm{RBS}(\mu,\delta)$, it is possible to show
that $b\,T \sim \textrm{RBS}(b\,\mu, \delta)$, with $b>0$,
$1/T \sim \textrm{RBS}(1/\beta, \delta)$
and $a(T)\sim \textrm{N}(0,1)$. On the other hand, if $Z\sim \textrm{N}(0,1)$,
since the function $t\mapsto a(t)$
is strictly increasing, it follows that
$a^{-\perp}(Z)\sim \textrm{RBS}(\mu,\delta).$
\item
Let $T_1\sim \textrm{RBS}(\mu_1,\delta_1)$ and $T_2\sim\textrm{RBS}(\mu_2,\delta_2)$
be two RVs statistically independent such that $T_1$ represents
``strength'' and $T_2$ represents ``stress'',
then the reliability of a component is given by
$R=\textrm{P}(T_1<T_2).$ Since
$a_k(T_k)\sim \textrm{N}(0,1)$, with
$a_k(t) = a(t;\alpha_{k},\beta_{k})$,
$\alpha_k=\sqrt{2/\delta_k}$ and $\beta_k=\mu_k\delta_k/(\delta_k+1)$ for each $k=1,2$,
by the independence of the RVs, we have that
\[
{Z_*}
=
\frac{\alpha_{1} \,a_1(T_1)-\alpha_{2} \,a_2(T_2)}{\sqrt{\alpha_{1}^2+\alpha_{2}^2}}\sim \textrm{N}(0,1).
\]
Then, $R= \textrm{P}({Z_*}<0)=1/2$ whenever $\beta_{1}=\beta_{2}$.
\item
Let $U\sim\chi_3^2$, where $\chi_3^2$ denotes the chi-squared distribution with
$3$ degrees of freedom, and
$g(\cdot;\mu, \delta)$ be the PDF of the RV $a^{-\perp}(\sqrt{U})$.
Then,
\[
g(u;\mu, \delta)
=
2\, a^2(u)
\,
f(u;\mu, \delta), \quad u>0.
\]
\end{enumerate}

\section{The bivariate RBS distribution}\label{sec:3}

\subsection{Density and shape analysis}

The bivariate random vector $\bm{T}=(T_{1},T_{2})^{\top}$ is said to follow a BRBS
distribution with parameters $\mu_{1}$, $\mu_{2}$, $\delta_{1}$, $\delta_{2}$, $\rho$,
denoted by $\bm{T}\sim\textrm{BRBS}(\mu_{1},\mu_{2},\delta_{1},\delta_{2},\rho)$, if
the joint CDF of $T_{1}$ and $T_{2}$ can be expressed as
\begin{eqnarray}\label{sec2:01}
F_{T_1,T_2}(t_1,t_2)
&=
\textrm{P}(T_{1}\leq{t}_{1},T_{2}\leq{t}_{2})
&=
\Phi_{2}
\left(
\sqrt{\frac{\delta_{1}}{2}}
\left(a_{1} - b_{1} \right),\sqrt{\frac{\delta_{2}}{2}}\left( a_{2}-b_{2} \right);\rho
\right),
\end{eqnarray}
where
\[
a_{k}=\sqrt{\frac{\left(\delta _k+1\right) t_{k}}{\delta_k \mu _k}},
\quad
b_{k}=\sqrt{\frac{\delta _k \mu _k}{\left(\delta _k+1\right) t_{k}}},
\quad
k=1,2,
\]
${t}_{1}>{0}$, ${t}_{2}>{0}$, $\mu_{1}>0$, $\delta_{1}>0$,
$\mu_{2}>0$, $\delta_{2}>0$, $|\rho|<1$, and $\Phi_{2}(u,v;\rho)$ is the
standard bivariate normal CDF with correlation coefficient $\rho$.
It follows that the joint PDF associated with \eqref{sec2:01} is given by
\begin{eqnarray}\label{sec2:02}
f_{T_{1},T_{2}}(t_{1},t_{2})
&= \displaystyle
\phi_{2}\left( \sqrt{\frac{\delta_{1}}{2}}\left( a_{1}-b_{1} \right),
\sqrt{\frac{\delta_{2}}{2}}\left( a_{2} - b_{2}\right);\rho \right)
\prod_{k=1}^{2}
\frac{\sqrt{\delta_k}}{2 \sqrt{2} t_k} \left(a_{k}+b_{k}\right),
\end{eqnarray}
where $\phi_{2}(u,v;\rho)$ is a normal joint PDF given by
$$
\phi_{2}(u,v;\rho)
=
\frac{1}{2\pi\sqrt{1-\rho^2}} \
\exp\left(-\frac{1}{2} Q(u,v)\right),
\quad \textrm{with} \
Q(u,v)=\frac{1}{(1-\rho^2)} (u^2-2\rho{u}{v}+v^2), \
u,v\in\mathbb{R}.
$$
Following the notation in \eqref{skewe-function} and considering  $a_k(t) =  a(t;\alpha_{k},\beta_{k})$,
with $\alpha_{k}=\sqrt{2/\delta_k}$ and $\beta_{k}=\mu_k\delta_k/(\delta_k+1)$, for $k=1,2$,
note that
\begin{eqnarray*}
F_{T_1,T_2}(t_1,t_2)
=
\Phi_2\left(a_1(t_1),a_2(t_2);\rho\right),
\quad 
f_{T_{1},T_{2}}(t_{1},t_{2})
=
\phi_2\left(a_1(t_1),a_2(t_2);\rho\right) \, \prod_{k=1}^{2}a_k'(t_k),
\end{eqnarray*}
where $a_k'(t_k)=\frac{\partial}{\partial t_k}a_k(t_k)$, for $k=1,2.$

\begin{figure}[h!!!]
\centering
\vspace{-0.6cm}
\psfrag{f}[c][c]{\tiny{$f_{T_{1},T_{2}}(t_{1},t_{2})$}}
\psfrag{h}[c][c]{\tiny{$h_{T_{1},T_{2}}(t_{1},t_{2})$}}
\psfrag{a}[c][c]{\tiny{$t_{1}$}}
\psfrag{b}[c][c]{\tiny{$t_{2}$}}
\psfrag{0}[c][c]{\tiny{0}}
\psfrag{1}[c][c]{\tiny{1}}
\psfrag{2}[c][c]{\tiny{2}}
\psfrag{3}[c][c]{\tiny{3}}
\psfrag{4}[c][c]{\tiny{4}}
\psfrag{5}[c][c]{\tiny{5}}
\psfrag{6}[c][c]{\tiny{6}}
\psfrag{7}[c][c]{\tiny{7}}
\psfrag{0.00}[c][c]{\tiny{0.0}}
\psfrag{0.000}[c][c]{\tiny{0.0}}
\psfrag{0.02}[c][c]{\tiny{0.02}}
\psfrag{0.04}[c][c]{\tiny{0.04}}
\psfrag{0.005}[c][c]{\tiny{.005}}
\psfrag{0.010}[c][c]{\tiny{.010}}
\psfrag{0.015}[c][c]{\tiny{.015}}
\psfrag{0.020}[c][c]{\tiny{.020}}
\psfrag{0.06}[c][c]{\tiny{0.06}}
\psfrag{0.08}[c][c]{\tiny{0.08}}
\psfrag{0.10}[c][c]{\tiny{0.10}}
\psfrag{0.12}[c][c]{\tiny{0.12}}
\psfrag{1.00}[c][c]{\tiny{1.00}}
\psfrag{0.0}[c][c]{\tiny{0.0}}
\psfrag{0.1}[c][c]{\tiny{0.1}}
\psfrag{0.2}[c][c]{\tiny{0.2}}
\psfrag{0.3}[c][c]{\tiny{0.3}}
\psfrag{0.4}[c][c]{\tiny{0.4}}
\psfrag{0.5}[c][c]{\tiny{0.5}}
\psfrag{0.6}[c][c]{\tiny{0.6}}
\psfrag{0.7}[c][c]{\tiny{0.7}}
\psfrag{0.8}[c][c]{\tiny{0.8}}
\psfrag{0.9}[c][c]{\tiny{0.9}}
\psfrag{1.0}[c][c]{\tiny{1.0}}
\psfrag{1.5}[c][c]{\tiny{1.5}}
\psfrag{2.0}[c][c]{\tiny{2.0}}
\psfrag{2.5}[c][c]{\tiny{2.5}}
\psfrag{3.0}[c][c]{\tiny{3.0}}
{\includegraphics[height=7.0cm,width=7.0cm,angle=-90]{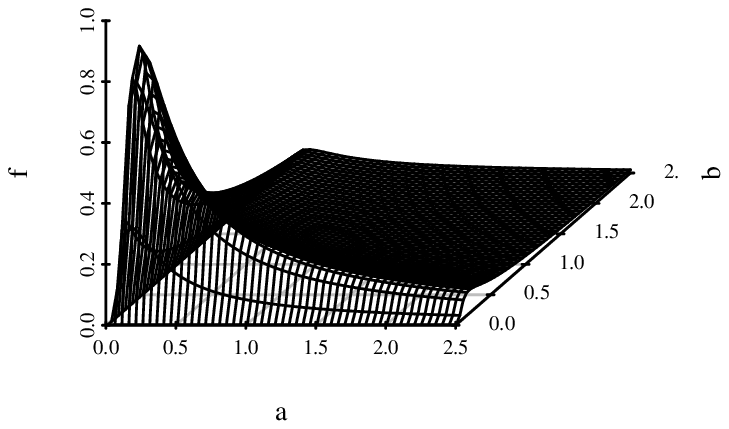}}
{\includegraphics[height=7.0cm,width=7.0cm,angle=-90]{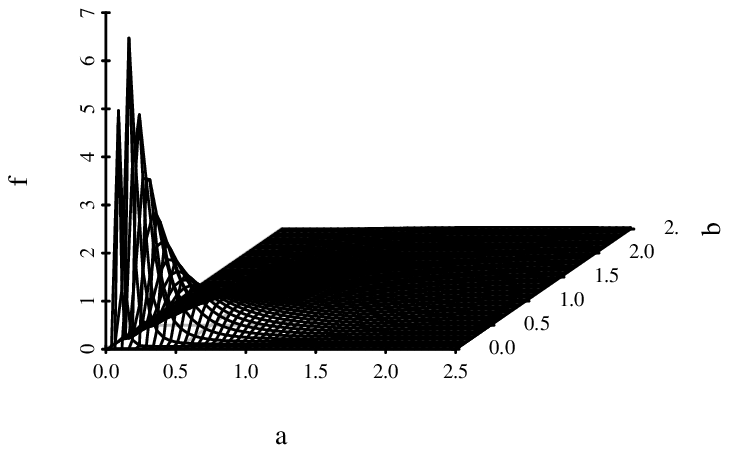}}\\[-2cm]
{\includegraphics[height=7.0cm,width=7.0cm,angle=-90]{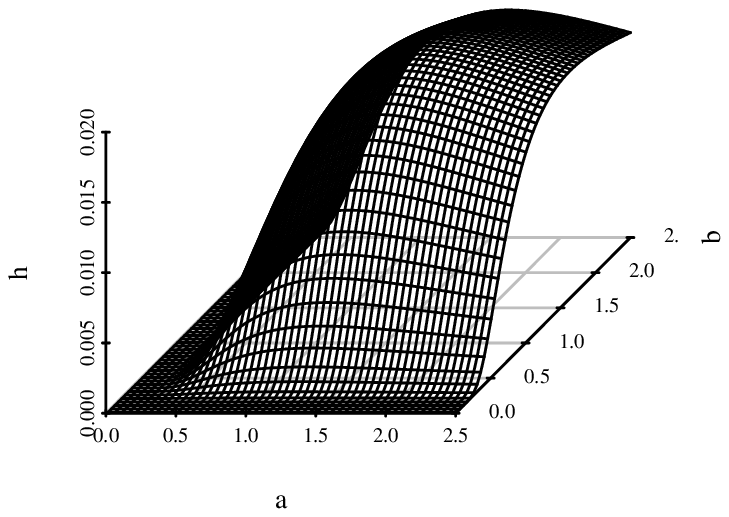}}
{\includegraphics[height=7.0cm,width=7.0cm,angle=-90]{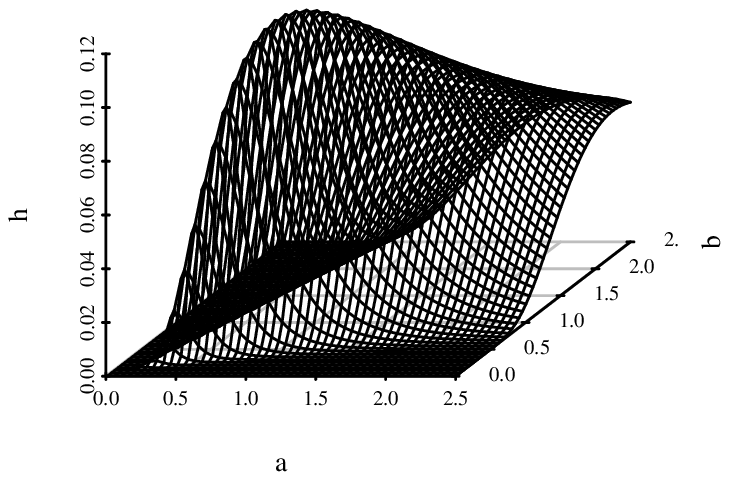}}\\
\vspace{-0.6cm}
\caption{\small The joint PDFs (top) of $(T_{1},T_{2})$ when $\mu_1=\mu_2=2$, $\delta_1=\delta_2=1$ and $\rho=0.1$ (left),
$\rho=0.9$ (right); and the joint HRs (bottom) of $(T_{1},T_{2})$ when $\mu_1=\mu_2=10$, $\delta_1=\delta_2=1.5$ and $\rho=0.1$ (left),
$\rho=0.9$ (right).
}
\label{fig:bivs}
\end{figure}

From now on, we will use the following notation
\begin{align}\label{cp}
c_{j,k}(t,w;\rho) = \frac{1}{\sqrt{1-\rho^2}}\, \big[a_j(t)-\rho a_k(w)\big],
\quad t>0, w>0,
\end{align}
for $j,k=1,2$.
\begin{Lemma}\label{pre-lemma}
Some important properties of the function $c_{j,k}(t,w;\rho)$ are:
\begin{enumerate}
\item $t\mapsto c_{j,k}(t,w;\rho)$ is an increasing function for all $w$.

If $\rho<0$ ($>0$), the function $w\mapsto c_{j,k}(t,w;\rho)$ is increasing (decreasing)
for all $t$.
\item $c_{1,2}(t,w;\rho)\geqslant 1$ whenever
$t\geqslant a_1^{-\perp}\big(\sqrt{1-\rho^2}\, (1+\rho)\big)$,
$w>a_2^{-\perp}(\sqrt{1-\rho^2})$ and $\rho<0$.
\item $c_{1,2}(t,w;\rho)\geqslant 1$ whenever
$t\geqslant a_1^{-\perp}\big(\sqrt{1-\rho^2}\big)$,
$w<\beta_2$ $(>\beta_2)$ and $\rho>0$ $(<0)$.
\item Let $\rho<0$.
For $v>w\geqslant a_2^{-\perp}(\sqrt{1-\rho^2})$ and
$t\geqslant a_1^{-\perp}\big(\sqrt{1-\rho^2}\, (1+\rho)\big)$, we have
\[
\left[\frac{1}{c_{1,2}(t,v;\rho)}-\frac{1}{c^3_{1,2}(t,v;\rho)}\right]
c_{1,2}(t,w;\rho)\geqslant 1.
\]
\item Let $\rho>0$ $(<0)$.
For $v>t\geqslant a_1^{-\perp}(\sqrt{1-\rho^2})$ and $w<\beta_{2}$ $(>\beta_2)$, we have
\[
\left[\frac{1}{c_{1,2}(t,w;\rho)}-\frac{1}{c^3_{1,2}(t,w;\rho)}\right]
c_{1,2}(v,w;\rho)\geqslant 1.
\]
\end{enumerate}
\end{Lemma}
The joint PDFs and HRs of $(T_{1},T_{2})$ are unimodal and the surface plots for
some values of the parameters are presented
in Figure~\ref{fig:bivs}.

Next, some results on the unimodality properties of \textrm{BRBS} distribution are obtained. We will consider the following hypothesis:
\\
\noindent
{\bf Hypothesis 1.}
Let $\bm{T}=(T_{1},T_{2})^{\top}\sim\textrm{BRBS}(\mu_{1},\mu_{2},\delta_{1},\delta_{2},\rho)$,
$\alpha_{k}=\sqrt{2/\delta_k}$ and $\beta_{k}=\mu_k\delta_k/(\delta_k+1)$, for $k=1,2$.
\begin{enumerate}
\item $\rho<\min\left\{\frac{\alpha_2}{\alpha_1},\, \frac{\alpha_1}{\alpha_2}\right\}$,
\item $\alpha_1>\frac{\tau_0}{(1-\rho^2)}
\max\left\{(\frac{1}{\alpha_1}-\rho\frac{1}{\alpha_2}),\,
(\frac{1}{\alpha_2}-\rho\frac{1}{\alpha_1})\right\}$,
\quad \textrm{where} \ $\tau_0\approx 96/10$.
\end{enumerate}
\begin{Proposition}\label{prop-del}
Under Hypothesis 1 there is an unique constant $c>0$ such that the point
$(\widetilde{t}_1,\widetilde{t}_2)=(c\beta_1,c\beta_2)$ is critical for $f_{T_1,T_2}.$
\end{Proposition}

\begin{Theorem}[Unimodality]\label{theo:1}
Under Hypothesis 1 there is an unique constant $c>0$ in the interval $(0,2\sqrt{3}-3)$
such that the point $(\widetilde{t}_1,\widetilde{t}_2)=(c\beta_1,c\beta_2)$ is a mode
for $f_{T_1,T_2}.$ That is, under Hypothesis 1 the \textrm{BRBS} distribution is unimodal.
\end{Theorem}

\subsection{Properties of the \textrm{BRBS} distribution}
\begin{Proposition}[Marginal functions]\label{Marginal}
Let $\bm{T}=(T_{1},T_{2})^{\top}\sim\textrm{BRBS}(\mu_{1},\mu_{2},\delta_{1},\delta_{2},\rho)$.
The marginal PDFs, denoted $f_{T_k}(\cdot)$, and the marginal CDFs, denoted $F_{T_k}(\cdot)$,
are given by
\[
f_{T_k}(t_k)
=
f(t_k;\mu_k,\delta_k),
\quad
F_{T_k}(t_k)=F(t_k;\mu_k,\delta_k),
\quad k=1,2,
\]
respectively,
where $f(\cdot;\mu_k,\delta_k)$ and $F(\cdot;\mu_k,\delta_k)$ are the PDF and the \textrm{CDF}
of the \textrm{RBS} distribution defined in \eqref{eq01}.
That is, $T_k\sim\textrm{RBS}(\mu_k,\delta_k)$.
\end{Proposition}

\begin{Proposition}[Reliability function]\label{Reliability function}
If\,\,$\bm{T}=(T_{1},T_{2})^{\top}\sim\textrm{BRBS}(\mu_{1},\mu_{2},\delta_{1},\delta_{2},\rho)$,
the reliability
$R=\textrm{P}(T_1<T_2)$
can be expressed as
\begin{align*}
R
=
\textrm{E}
\left[
\Phi\big(c_{2,1}(T_1,T_1;\rho)\big)
\right],
\quad \ T_1\sim\textrm{RBS}(\mu_1,\delta_1),
\end{align*}
where $c_{2,1}(s,t;\rho)$ is defined in \eqref{cp}.
\end{Proposition}

\begin{Remark}
If $T_1$ and $T_2$ are identically distributed RVs following a RBS distribution, then
\[
\begin{array}{c}
R
=
\textrm{E}\left[\Phi\left(\sqrt{\frac{1-\rho}{1+\rho}} \ Z\right) \right],
\end{array}
\]
where $Z\sim \textrm{N}(0,1)$, and if in addition $\rho=0$, then
$R
=
1/2.
$
\end{Remark}
\begin{Proposition}\label{prop-pre}
Let $\bm{T}=(T_{1},T_{2})^{\top}\sim\textrm{BRBS}(\mu_{1},\mu_{2},\delta_{1},\delta_{2},\rho)$.
Then,
\begin{enumerate}
\item
$
\int_{t_2}^{\infty} \int_{t_1}^{\infty} f_{T_1,T_2}(u,v) \,\mathrm{d}u \, \mathrm{d}v
=
\int_{t_2}^{\infty} f(v;\mu_2,\delta_2)\big\{1-\Phi\big(c_{1,2}(t_1,v;\rho)\big)\big\} \, \mathrm{d}v
$;
\item
$
\int_{t_1}^{\infty}\int_{v}^{\infty}f_{T_1,T_2}(u,t_2) \,\mathrm{d}u \, \mathrm{d}v
=
f(t_2;\mu_2,\delta_2) \,
\int_{t_1}^{\infty}
1-\Phi\big(c_{1,2}(v,t_2;\rho)\big) \, \mathrm{d}v,
$
\end{enumerate}
where $f(t;\mu_2,\delta_2)$ and $c_{1,2}(v,t_2;\rho)$ are defined in
\eqref{eq01} and  \eqref{cp}, respectively.
\end{Proposition}

Some authors like \cite{basu:71} or \cite{puri:74}
define the multivariate \textrm{HR} as a scalar quantity.
In the bivariate case, Basu gives the \textrm{HR} as
\[
h_{T_1,T_2}(t_1,t_2)
=
\frac{f_{T_1,T_2}(t_1,t_2)}{S(t_1,t_2)},
\quad \textrm{where} \quad
S(t_1,t_2)=\int_{t_1}^{\infty}\int_{t_2}^{\infty} f_{T_1,T_2}(w,t)\, \textrm{d}w \, \textrm{d}t
\]
is the bivariate \textrm{SF}. An analogous procedure to Proposition \ref{prop-pre} shows that
\begin{eqnarray}\label{sf}
S(t_1,t_2)
=
\int_{t_1}^\infty f(w;\mu_1,\delta_1)\left\{1-\Phi\big(c_{2,1}(t_2,w;\rho)\big) \right\}\,\textrm{d}w.
\end{eqnarray}
Therefore, we get
\begin{align}\label{exp-hr}
h_{T_1,T_2}(t_1,t_2)
=
\frac{f(t_1;\mu_1,\delta_1) \phi\big(c_{2,1}(t_2,t_1;\rho)\big)}
{\int_{t_1}^\infty f(w;\mu_1,\delta_1)\left\{1-\Phi\big(c_{2,1}(t_2,w;\rho)\big) \right\}\,\textrm{d}w}.
\end{align}
%
%
\begin{Proposition}\label{prop:pre}
The function  
$G(t_1,t_2)=\int_{t_1}^\infty f(w;\mu_1,\delta_1)\left\{1-\Phi\big(c_{2,1}(t_2,w;\rho)\big) \right\}\,\mathrm{d}w$	is decreasing in both $t_1$ and $t_2$.
\end{Proposition}
\begin{Proposition}\label{prop:3.5}
For the \textrm{BRBS} distribution, 
\begin{enumerate}
\item if $\rho<0$, $t_1\leqslant \beta_1$ and $t_2\leqslant \beta_2$;
then the function $h_{T_1,T_2}(t_1,t_2)$ is increasing in $t_1$.
\item if  $\rho> 0$, $t_1\geqslant \beta_1$ and $t_2\leqslant \beta_2$;
then the function $h_{T_1,T_2}(t_1,t_2)$ is increasing in $t_2$.
\item if  $\rho= 0$ and $t_2\leqslant \beta_2$;
then the function $h_{T_1,T_2}(t_1,t_2)$ is increasing in $t_2$.
\end{enumerate}
\end{Proposition}
\begin{Proposition}\label{prop:3.6}
Let $\bm{T}=(T_{1},T_{2})^{\top}\sim\textrm{BRBS}(\mu_{1},\mu_{2},\delta_{1},\delta_{2},\rho)$
and $R=\textrm{P}(T_1<T_2)$ be the reliability. Then,
\[
\int_{0}^{\infty} h_{T_1,T_2}(t_1,t_2=t_1)\, \mathrm{d}t_1
\geqslant \frac{R}{1-R}.
\]
\end{Proposition}

\begin{Proposition}\label{prop-int}
Let $X\sim \textrm{N}(b,\sigma^2)$, $b\in\mathbb{R}$ and $\sigma>0$.
Then,
\[
\begin{array}{c}
\textrm{E}[a^{-\perp}(X)]=\frac{\beta\alpha^2}{2}(\sigma^2+b^2)+\beta.
\end{array}
\]
\end{Proposition}

Let $X$ and $Y$ be RVs, the covariance and correlation of $X$ and $Y$,
as usual, are denoted by $\textrm{Cov}(X,Y)$ and  $\rho(X,Y)$ respectively.
The following result tells us that two RVs with \textrm{BRBS} distribution
are associated and correlated positively.
\begin{Theorem}\label{theorem-p}
Let $\bm{T}=(T_{1},T_{2})^{\top}\sim\textrm{BRBS}(\mu_{1},\mu_{2},\delta_{1},\delta_{2},\rho)$,
$\alpha_{k}=\sqrt{2/\delta_k}$ and $\beta_{k}=\mu_k\delta_k/(\delta_k+1)$, for $k=1,2$.
Then,
\begin{enumerate}
\item
$
\textrm{E}(T_1|T_2=t_2)
=
\frac{\beta_1\alpha_1^2}{2}
\left[
\frac{2}{\alpha_1^2}+ (1-\rho^2)+ \rho^2 a_2^2(t_2)
\right],
$
\item
$
\textrm{E}(T_1 T_2)
=
\frac{\beta_1\beta_2}{4}
\big[
4
+
2(\alpha_1^2 + \alpha_2^2)
+
{\alpha_1^2\alpha_2^2}(1+2\rho^2)
\big],
$
\item
$
\textrm{Cov}(T_1,T_2)=\frac{\rho^2}{2}\prod_{k=1}^{2}{\beta_k\alpha_k^2},
$
\item
$
\rho(T_1,T_2)
=
2\rho^2
\prod_{k=1}^{2}
\frac{\alpha_k}{\sqrt{4+5\alpha_k^2}}
$
,
$
|\rho|< \frac{1}{\sqrt{2}}
\prod_{k=1}^{2}\frac{\sqrt[4]{4+5\alpha_k^2}}{\sqrt{\alpha_k}}.
$
\end{enumerate}
\end{Theorem}

Let $(X,Y)$ be a bivariate positive RV with bivariate $\textrm{SF}$ $S(x,y)$
(defined in \eqref{sf}) and with $\textrm{E}[XY]$ finite.
If we assume that the sampling probability of $(X,Y)$ is proportional to $XY$,
the recurrence times are $(X^{eq}, Y^{eq})$, where $X^{eq}=X^{sb}U_1$
and $Y^{eq} = Y^{sb}U_2$, where $U_1$ and $U_2$ have independent
uniform distributions in $(0,1)$ and are independent of $(X^{sb},Y^{sb})$. The vector
$(X^{sb}, Y^{sb})$ is known as the size biased random vector and has PDF defined by
\begin{align*}
f^{sb}(x,y)=\frac{xy}{\textrm{E}[XY]}\, f(x,y).
\end{align*}
Under the previous hypothesis, the joint PDF of the bivariate equilibrium distribution
$(X^{eq}, Y^{eq})$ is given by (see \cite{nav06} or \cite{sar10})
\begin{align}\label{equilibrium}
f^{eq}(x,y)
=
\frac{S(x,y)}
{\textrm{E}[XY]}, \quad x> 0, y> 0.
\end{align}
\begin{Proposition}[Equilibrium distribution]\label{Equilibrium distribution}
According to \eqref{equilibrium}, the equilibrium PDF associated with \eqref{eq01} is
\[
f^{eq}(t_1,t_2)
=
\frac{\int_{t_1}^\infty f(w;\mu_1,\delta_1)\left\{1-\Phi\big(c_{2,1}(t_2,w;\rho)\big) \right\}
\,\mathrm{d}w}
{\frac{\beta_1\beta_2}{4}
\big[
4
+
2(\alpha_1^2 + \alpha_2^2)
+
{\alpha_1^2\alpha_2^2}(1+2\rho^2)
\big]}.
\]
\end{Proposition}
\begin{Proposition}\label{prop:3.9}
For the \textrm{BRBS} distribution,
the function $f^{eq}(t_1,t_2)$ is decreasing in both $t_1$ and $t_2$.
\end{Proposition}
\begin{Theorem}\label{theo:3.3}
Let $\bm{T}=(T_{1},T_{2})^{\top}\sim\textrm{BRBS}(\mu_{1},\mu_{2},\delta_{1},\delta_{2},\rho)$
and
$\beta_{k}=\mu_k\delta_k/(\delta_k+1)$, for $k=1,2$.
Then,
\begin{enumerate}
\item
$
\bm{T}^{-1}=(T_{1}^{-1},T_{2}^{-1})^{\top}\sim
\textrm{BRBS}
\left(
{\mu_1/\beta_{1}^2},
{\mu_2/\beta_{2}^2},
\delta_{2},\rho
\right)
$;
\item
$
\bm{T}_{1}^{-1}=(T_{1}^{-1},T_{2})^{\top}\sim
\textrm{BRBS}
\left(
{\mu_1/\beta_{1}^2},
\delta_{1},
\mu_{2},
\delta_{2},
-\rho
\right)$;
\item
$\bm{T}_{2}^{-1}=(T_{1},T_{2}^{-1})^{\top}\sim
\textrm{BRBS}
\left(
\mu_{1},
\delta_{1},
{\mu_2/\beta_{2}^2},
\delta_{2},
-\rho
\right)$.
\end{enumerate}
\end{Theorem}

\subsection{Failure rate with presence of dependence}

A real function $K(x,y)$, which is defined on $\mathcal{X}$ and $\mathcal{Y}$,
$\mathcal{X}$ and $\mathcal{Y}$ are linearly ordered sets, is said to be
totally positive of order two ($\textrm{TP}_2$) and reverse rule of order two ($\textrm{RR}_2$) in $x\in \mathcal{X}$
and $y\in \mathcal{Y}$ if
\begin{align*}
K(x,y)\geqslant 0 \ (\leqslant 0), \quad x\in \mathcal{X}, \ y\in \mathcal{Y},
\end{align*}
\begin{align*}
K(x,y)K(x',y')\geqslant K(x,y')K(x',y),\quad \textrm{for all} \ \  x<x'\in \mathcal{X}, \quad y<y'\in \mathcal{Y}.
\end{align*}

We also define a local dependence function (\textrm{LDF})
\[
\gamma_K(x,y) = {\partial^2\over\partial x \partial y}\log K(x,y).
\]
The \textrm{LDF}, $\gamma_K(x,y)$, can be defined for any positive and mixed
differentiable function $K(x,y)$, which does not need to be a density function.
See \cite{hollandw:87} for definition and properties of the local dependence function.

\begin{Theorem}\citep{hollandw:87}\label{Holland and Wang}
A function $K(x,y)$ is $\textrm{TP}_2$ $(\textrm{RR}_2)$ if and only if $\gamma_K(x,y)\geqslant 0$
$(\leqslant 0)$.
\end{Theorem}
It can be verified that
\begin{align}\label{tp2}
\gamma_{f_{T_{1},T_{2}}}(t_1,t_2)=\left(\rho\over 1-\rho^2\right) \,
\prod_{k=1}^{2} a'_k(t_k)>0 \ (<0),
\quad t_1>0, t_2>0,
\end{align}
whenever $\rho>0$ $(<0)$.
Then, by Theorem \ref{Holland and Wang} the joint PDF $f_{T_{1},T_{2}}(t_1,t_2)$
is an example of a function $\textrm{TP}_2$ $(\textrm{RR}_2)$ when $\rho>0$ $(<0)$.
The \textrm{LDF} will be used for studying the monotonicity of certain HRs.
If $(X,Y)^{\top}$ is a bivariate random vector distributed according to
$F(x,y)$, then one can consider $F(x|Y\in A)$ (the \textrm{CDF} of $X$ given that $Y\in A$).
The  conditional \textrm{HR} (\textrm{CHR}) of $X$ given $Y\in A$ and
mean residual function (\textrm{MRF}),
using obvious notation, are defined by
\[
h(x|Y\in A)
=
{f(x|Y\in A)\over S(x|Y\in A)},
\quad
m(x|Y\in A)
=
{\int_{x}^{\infty}S(t|Y\in A) \,\textrm{d}t \over S(x|Y\in A)},
\]
respectively, where
\[
S(x|Y\in A) =  1-F(x|Y\in A)
\]
denotes the conditional SF (\textrm{CSF}).

We know a lot about the normal distribution. For example, if $Z$ has
a $\textrm{N}(0,1)$ distribution we have (\cite{book:feller}, Section 7.1)
\begin{align}\label{The Gaussian Tail Inequality}
\left(\frac{1}{z}-\frac{1}{z^3}\right)\frac{\exp(-z^2/2)}{\sqrt{2\pi}}
\leqslant
\textrm{P}(Z>z)
\leqslant
\frac{1}{z}\,\frac{\exp(-z^2/2)}{\sqrt{2\pi}},
\quad
\forall z>0.
\end{align}
Clearly the inequalities are useful only for larger $z$, because as $z$ decreases to zero
the lower bound goes to $-\infty$ and the upper bound goes to $+\infty$.
The inequality above is known as the Gaussian tail inequality.
\begin{Theorem}[Monotonicity]\label{prop-failures}
Let us assume that $\bm{T}=(T_{1},T_{2})^{\top}$ has a
$\textrm{BRBS}(\mu_{1},\mu_{2},\delta_{1},\delta_{2},\rho)$ distribution and  consider
$\alpha_{k}=\sqrt{2/\delta_k}$ and
$\beta_{k}=\mu_k\delta_k/(\delta_k+1)$ for $k=1,2$. Then,
\begin{enumerate}
\item The \textrm{CHR}:  $t_2\mapsto h(t_1|T_2>t_2)$ is decreasing for all
$t_2>a_2^{-\perp}(\sqrt{1-\rho^2})$ and for all
$t_1\geqslant a_1^{-\perp}\big(\sqrt{1-\rho^2}\, (1+\rho)\big)$, whenever $\rho<0$.
\item
The \textrm{MRF}:
$t_2\mapsto m(t_1|T_2=t_2)$ is increasing for all $t_2<\beta_{2}$ and for all
$t_1\geqslant a_1^{-\perp}(\sqrt{1-\rho^2})$, whenever $\rho>0$.
\item The \textrm{CSF} :  $t_2\mapsto S(t_1|T_2>t_2)$ is increasing for all
$t_2>a_2^{-\perp}(\sqrt{1-\rho^2})$ and for all
$t_1\geqslant a_1^{-\perp}\big(\sqrt{1-\rho^2}\, (1+\rho)\big)$, whenever $\rho<0$.
\item The \textrm{CHR}: $t_2\mapsto h(t_1|T_2=t_2)$ is decreasing (increasing) when
$\rho>0$ $(<0)$.
\end{enumerate}
\end{Theorem}

\section{Estimation and inference}\label{sec:4}

\subsection{Maximum likelihood estimation}

\noindent Let $\{(t_{1i},t_{2i}),i=1,\ldots,n\}$ be a bivariate random sample
of size $n$ from the $\textrm{BRBS}(\mu_{1},\mu_{2},\delta_{1},\delta_{2},\rho)$ distribution with
PDF as given in \eqref{sec2:02}. Then, the log-likelihood function,
without the additive constant, is given by
%
\begin{align}\label{sec2.1:01}
\hspace*{-0,3cm}
\ell({\bm\theta})=&
-
\frac{n}{2}\log(1-\rho ^2)
+
\frac{1}{4 \left(1-\rho ^2\right)}
\sum_{i=1}^{n}
\left\{
2\rho \prod_{k=1}^2 \sqrt{\delta _k}\left(a_{ki}-b_{ki}\right)  \right.  \nonumber
\\&
+\left.
\sum_{k=1}^{2}
\left(
\frac{\delta_k^2 \mu _k}{\left(\delta_k+1\right) t_{ki}}
-
2 \delta_k a_{ki} b_{ki}
+
\frac{\left(\delta_k+1\right) t_{ki}}{\mu _k}
\right)
\right\}     
+
\sum_{k=1}^{2}
\left(
\frac{n}{2} \log(\delta_k)
+
\sum_{i=1}^{n} \log\left(a_{ki}+b_{ki}\right)
\right),
\end{align}
where ${\bm\theta}=(\mu_{1},\mu_{2},\delta_{1},\delta_{2},\rho)^{\top}$,
\[
a_{ki}=\sqrt{\frac{\left(\delta _k+1\right) t_{ki}}{\delta _k \mu _k}},
b_{ki}=\sqrt{\frac{\delta _k \mu _k}{\left(\delta _k+1\right) t_{ki}}}, \quad
k=1,2.
\]
Note that
$\big((a_{1}-b_{1})\sqrt{{\delta_{1}}/{2}}, (a_{2}-b_{2})\sqrt{{\delta_{2}}/{2}}\big)$
is bivariate normal distributed with mean vector $(0,0)^{\top}$ and covariance matrix
$
\begin{pmatrix}
1 & \rho \\
\rho & 1
\end{pmatrix}.
$
From this result, we see that, for given $\mu_{1}$,
$\mu_{2}$, $\delta_{1}$, $\delta_{2}$, the ML estimator of $\rho$ is
\begin{align}\label{sec2.1:02}
\widehat{\rho}(\mu_{1},\mu_{2},\delta_{1},\delta_{2})
=
\frac{
\sum_{i=1}^{n}
\prod_{k=1}^{2}
\sqrt{\frac{\widehat{\delta}_{k}}{2}}
\left( \widehat{a}_{ki}-\widehat{b}_{ki} \right)
}
{
\prod_{k=1}^{2}
\sqrt{
\sum_{i=1}^{n}
\left\{ \sqrt{\frac{\widehat{\delta}_{k}}{2}}\left( \widehat{a}_{ki}-\widehat{b}_{ki} \right)\right\}^2
}
},
\end{align}
where
\[
\widehat{a}_{ki}=\sqrt{\frac{\left(\widehat{\delta} _k+1\right) t_{ki}}{\widehat{\delta} _k \widehat{\mu} _k}},
\quad
\widehat{b}_{ki}=\sqrt{\frac{\widehat{\delta} _k \widehat{\mu} _k}{\left(\widehat{\delta} _k+1\right) t_{ki}}}, \quad
k=1,2.
\]

Therefore, when the parameters $\mu_{1}$, $\mu_{2}$, $\delta_{1}$ and $\delta_{2}$
are unknown, the ML estimators of $\mu_{1}$, $\mu_{2}$, $\delta_{1}$ and $\delta_{2}$ can be
obtained by maximizing the profile log-likelihood function
%
\begin{eqnarray}\label{sec2.1:03}
\ell_{p}({\bm\eta})
&=&
-\frac{n}{2}\log(1-\widehat{\rho}(\mu_{1},\mu_{2},\delta_{1},\delta_{2}) ^2)
+
\sum_{k=1}^{2}
\left(
\frac{n}{2}\log(\delta_k)
+
\sum_{i=1}^{n}\log\left(a_{ki}+b_{ki}\right)
\right)
\nonumber
\\&&
-
\frac{1}{4 \left(1-\widehat{\rho}(\mu_{1},\mu_{2},\delta_{1},\delta_{2}) ^2\right)}
\sum_{i=1}^{n}
\left\{
-2\widehat{\rho}(\mu_{1},\mu_{2},\delta_{1},\delta_{2})
\prod_{k=1}^2
\sqrt{\delta _k}
\left(a_{ki}-b_{ki}\right)
\right.
\nonumber
\\&&
\left.
+
\sum_{k=1}^{2}
\left(
\frac{\delta _k^2 \mu _k}{\left(\delta _k+1\right) t_{ki}}
-
2\delta _k a_{ki} b_{ki}
+
\frac{\left(\delta _k+1\right) t_{ki}}{\mu _k}
\right)
\right\},
\end{eqnarray}
where ${\bm\eta}=(\mu_{1},\mu_{2},\delta_{1},\delta_{2})^{\top}$. In order
to maximize function \eqref{sec2.1:03}
with respect to $\mu_{1}$, $\mu_{2}$, $\delta_{1}$, $\delta_{2}$,
one may use the Newton-Raphson algorithm or
some other optimization algorithm. Once $\widehat{\mu}_{1}$,
$\widehat{\delta}_{1}$, $\widehat{\mu}_{2}$ and $\widehat{\delta}_{2}$
are obtained, the ML estimators of $\rho$, say $\widehat{\rho}$, is
computed from \eqref{sec2.1:02}. Under some regularity
conditions \citep{ch:74},
the asymptotic distribution of $\widehat{\bm\theta}=(\widehat{\mu}_{1},\widehat{\mu}_{2},\widehat{\delta}_{1},\widehat{\delta}_{2},\widehat{\rho})$,
as $n\rightarrow{\infty}$, is given by
\begin{equation*}\label{sec2.1:04}
 \sqrt{n}(\widehat{\bm\theta}-{\bm\theta})\xrightarrow{\textrm{D}} {N}_{5}\left({\bm 0},\bm{J}^{-1}\right),
\end{equation*}
where $\xrightarrow{\textrm{D}}$ denotes convergence in distribution
and $\textrm{N}_{5}\left({\bm 0},\bm{J}^{-1}\right)$ denotes
a $5$-variate normal distribution with mean ${\bm 0}$ and
covariance matrix $\bm{J}^{-1}$. For the sake of space we
omit the elements of the matrix $\bm{J}$.

\subsection{Modified moment estimation}

\noindent Let $\{(t_{1i},t_{2i}),i=1,\ldots,n\}$ be a bivariate random sample of
size $n$ from $T\sim\textrm{BRBS}({\bm \theta})$, with ${\bm\theta}=(\mu_{1},\mu_{2},\delta_{1},\delta_{2},\rho)^{\top}$. 
Also, let the sample arithmetic and harmonic means be defined as
\begin{equation*}\label{sec2.2:01}
s_{k} = \frac{1}{n}\sum\limits_{i=1}^{n}t_{ki},
\quad
r_{k} = \left[\frac{1}{n}\sum\limits_{i=1}^{n}t_{ki}^{-1}\right]^{-1},\quad k=1,2,
\end{equation*}
respectively. Then, the MM estimators of $\mu_{1}$, $\delta_{1}$, $\mu_{2}$
and $\delta_{2}$ are obtained by equating
$\textrm{E}[T_{1}]$, $\textrm{E}[T_{1}^{-1}]$, $\textrm{E}[T_{2}]$ and $\textrm{E}[T^{-1}_{2}]$ to
the corresponding sample estimates, that is, $\textrm{E}[T_{k}] = s_{k}$ and $\textrm{E}[T_{k}^{-1}]=r_{k}^{-1}$, for $k=1,2$. Thus, we readily have
\begin{eqnarray}\label{sec2.2:03}
s_{k}
=
\widetilde{\mu}_{k},
\quad
r_{k}^{-1}
=
\frac{(\widetilde{\delta}_{k}+1)^2}{\widetilde{\mu}_{k}\widetilde{\delta}_{k}^2},
\quad k=1,2.
\end{eqnarray}
Solving \eqref{sec2.2:03} for $\mu_{1}$, $\mu_{2}$, $\delta_{1}$ and $\delta_{2}$,
we obtain the MM estimators of these parameters,
denoted by $\widetilde{\mu}_{1}$, $\widetilde{\delta}_{1}$, $\widetilde{\mu}_{2}$
and $\widetilde{\delta}_{2}$, namely,
\begin{equation*}\label{sec2.2:04}
\widetilde{\mu}_{k} =s_{k},
\quad
\widetilde{\delta}_{k}=\left[\sqrt{\frac{s_{k}}{r_{k}}}-1 \right]^{-1},
\quad k=1,2.
\end{equation*}
and then the MM estimator of $\rho$ as
\begin{align*}
\widetilde{\rho}(\mu_{1},\mu_{2},\delta_{1},\delta_{2})
=
\frac{
\sum_{i=1}^{n}
\prod_{k=1}^{2}
\sqrt{\frac{\widetilde{\delta}_{k}}{2}}
\left( \widetilde{a}_{ki}-\widetilde{b}_{ki} \right)
}
{
\prod_{k=1}^{2}
\sqrt{
\sum_{i=1}^{n}
\left\{ \sqrt{\frac{\widetilde{\delta}_{k}}{2}}\left( \widetilde{a}_{ki}-\widetilde{b}_{ki} \right)\right\}^2
}
},
\end{align*}
where
\[
\widetilde{a}_{ki}=\sqrt{\frac{\left(\widetilde{\delta} _k+1\right) t_{ki}}{\widetilde{\delta} _k \widetilde{\mu} _k}},
\quad
\widetilde{b}_{ki}=\sqrt{\frac{\widetilde{\delta} _k \widetilde{\mu} _k}{\left(\widetilde{\delta} _k+1\right) t_{ki}}}, \quad
k=1,2.
\]
Note that the MM estimators have explicit forms, then they can be used as the initial guess in the numerical procedure  for
computing the ML estimators.

\begin{Theorem}\label{theo:4.1}
The asymptotic distributions of $\widetilde{\mu}_{k}$ and $\widetilde{\delta}_{k}$, for $k=1,2$, are given by
\begin{equation*}\label{sec2.2:05}
\sqrt{n}(\widetilde{\mu}_{k}-{\mu}_{k})
\sim
\textrm{N}\left(0,\frac{\mu_{k}^2(2\delta_{k}+5)}{(\delta_{k}+1)^2} \right),
\quad
\sqrt{n}(\widetilde{\delta}_{k}-{\delta}_{k})\sim  \textrm{N}\left(0,2 \delta_{k}^2 \right)
\end{equation*}
respectively.
\end{Theorem}

\section{Numerical applications}\label{sec:5}

\subsection{Simulation study}

In this section, we carry out a simulation study
to evaluate the performance of the ML and MM estimators
of the BRBS model parameters. The simulation scenario considered the following:
the sample sizes $n \in \{10, 50, 100\}$; the values of the shape and scale parameters as
$\delta_{k}\in\{0.25,2.0\}$ and $\mu_{k}=2.0$, for $k=1,2$, respectively;
the values of $\rho$ are $0.00,0.25,0.50$ and $0.95$ (the results for
negative $\rho$ are quite similar so are omitted here); and $5,000$
MC replications. Note that the values of $\delta_{k}$ cover low and high skewness.
We also present the $90\%$ and $95\%$ probability coverages of
confidence intervals for the BRBS model.

Tables~\ref{tab:03}-\ref{tab:04} report the bias and mean squared error (MSE)
for the ML and MM estimates. A look at the results allows to conclude that, as $n$ increases,
the bias and MSE of all the estimators decrease,
tending to be unbiased, as expected.
Moreover, the performances of
the ML and MM estimators are quite similar in terms of bias and MSE.
Furthermore, we note that, as the values
of the shape parameters $\delta_{k}$ increase, the performances of the estimators of
$\mu_{k}$, the scale parameters, deteriorate.
In general, $\rho$ does not seem to have influence on the results.

\begin{table}[h!]
\scriptsize
\centering
\caption{Simulated values of biases and MSEs (within parentheses) of the
ML estimates in comparison with those of MM estimates ($\mu_{k} = 2.0$ and $\delta_{k}=0.25$, for $k=1,2$),
for the BRBS distribution.}\label{tab:03}
\vspace*{0,25cm}
\begin{tabular}{lllrrrrrrrrrr}
\hline\noalign{\smallskip}\\[-0.4cm]
$n$&  &&\multicolumn{5}{c}{ML} \\ \cline{4-8} \\[-0.3cm]
   & \multicolumn{1}{c}{$\rho$}&&\multicolumn{1}{c}{Bias($\widehat{\delta}_{1}$)}&\multicolumn{1}{c}{Bias($\widehat{\delta}_{2}$)}&
                      \multicolumn{1}{c}{Bias($\widehat{\mu}_{1}$)}&\multicolumn{1}{c}{Bias($\widehat{\mu}_{2}$)}&\multicolumn{1}{c}{Bias($\widehat{\rho}$)}\\
\noalign{\smallskip}\hline\noalign{\smallskip}
10       &0.00  &&0.0470\,(0.0687)   &0.0566\,(0.0650)   &  $-$0.0250\,(1.4560) & $-$0.0537\,(1.3561) &$-$0.0119\,(0.1181) \\
         &0.25  &&0.0492\,(0.0557)   &0.0549\,(0.0669)   &  $-$0.0884\,(1.4203) & $-$0.0899\,(1.3267) &$-$0.0057\,(0.1092)  \\
         &0.50  &&0.0487\,(0.0653)   &0.0507\,(0.0587)   &  $-$0.0430\,(1.3532) & $-$0.0218\,(1.4600) &$-$0.0082\,(0.0680)  \\
         &0.95  &&0.0480\,(0.0638)   &0.0467\,(0.0766)   &  $-$0.0374\,(1.2801) & $-$0.0265\,(1.2389) &$-$0.0077\,(0.0028)  \\[0.1cm]
50       &0.00  &&0.0073\,(0.0031)   &0.0105\,(0.0034)   &  $-$0.0157\,(0.2621) & $-$0.0209\,(0.2768) &$-$0.0071\,(0.0218) \\
         &0.25  &&0.0115\,(0.0035)   &0.0099\,(0.0032)   &  $-$0.0224\,(0.2560) & $-$0.0122\,(0.2622) &$-$0.0037\,(0.0178)  \\
         &0.50  &&0.0093\,(0.0035)   &0.0100\,(0.0035)   &  $-$0.0178\,(0.2748) & $-$0.0100\,(0.2743) &$-$0.0015\,(0.0118)  \\
         &0.95  &&0.0058\,(0.0033)   &0.0048\,(0.0031)   &  $-$0.0035\,(0.2255) & $-$0.0042\,(0.2320) &$-$0.0003\,(0.0002)  \\[0.1cm]
100      &0.00  &&0.0040\,(0.0015)   &0.0057\,(0.0016)   &  $-$0.0121\,(0.1385) & $-$0.0048\,(0.1360) &$-$0.0007\,(0.0105) \\
         &0.25  &&0.0026\,(0.0014)   &0.0053\,(0.0014)   &  $-$0.0067\,(0.1466) & $-$0.0220\,(0.1257) &$-$0.0004\,(0.0101)  \\
         &0.50  &&0.0053\,(0.0014)   &0.0016\,(0.0014)   &  $-$0.0159\,(0.1256) & $-$0.0100\,(0.1297) &$-$0.0019\,(0.0059)  \\
         &0.95  &&0.0038\,(0.0015)   &0.0036\,(0.0015)   &  $-$0.0043\,(0.1091) & $-$0.0067\,(0.1118) &$-$0.0001\,(0.0001)  \\[0.1cm]
\hline\noalign{\smallskip}
$n$&  &&\multicolumn{5}{c}{MM} \\ \cline{4-8} \\[-0.3cm]
   & \multicolumn{1}{c}{$\rho$}&&\multicolumn{1}{c}{Bias($\widetilde{\delta}_{1}$)}&\multicolumn{1}{c}{Bias($\widetilde{\delta}_{2}$)}&
                      \multicolumn{1}{c}{Bias($\widetilde{\mu}_{1}$)}&\multicolumn{1}{c}{Bias($\widetilde{\mu}_{2}$)}&\multicolumn{1}{c}{Bias($\widetilde{\rho}$)}\\
\noalign{\smallskip}\hline\noalign{\smallskip}
10       &0.00  &&0.0517\,(0.0695)   &0.0612\,(0.0658)   &  $-$0.0386\,(1.3952) & $-$0.0633\,(1.3487) &$-$0.0119\,(0.1103) \\
         &0.25  &&0.0541\,(0.0565)   &0.0598\,(0.0679)   &  $-$0.1187\,(1.3182) & $-$0.1224\,(1.2675) &$-$0.0021\,(0.1038)  \\
         &0.50  &&0.0554\,(0.0670)   &0.0575\,(0.0597)   &  $-$0.0435\,(1.3913) & $-$0.0602\,(1.5602) &$-$0.0228\,(0.0669)  \\
         &0.95  &&0.0595\,(0.0673)   &0.0583\,(0.0797)   &  $-$0.0422\,(1.3808) & $-$0.0230\,(1.3971) &$-$0.0122\,(0.0032)  \\[0.1cm]
50       &0.00  &&0.0080\,(0.0031)   &0.0112\,(0.0034)   &  $-$0.0164\,(0.2723) & $-$0.0299\,(0.2884) &$-$0.0068\,(0.0213) \\
         &0.25  &&0.0123\,(0.0035)   &0.0107\,(0.0032)   &  $-$0.0230\,(0.2692) & $-$0.0123\,(0.2724) &$-$0.0019\,(0.0175)  \\
         &0.50  &&0.0104\,(0.0035)   &0.0111\,(0.0035)   &  $-$0.0130\,(0.2971) & $-$0.0077\,(0.3001) &$-$0.0052\,(0.0118)  \\
         &0.95  &&0.0083\,(0.0034)   &0.0073\,(0.0032)   &  $-$0.0058\,(0.2657) & $-$0.0132\,(0.2684) &$-$0.0006\,(0.0002)  \\[0.1cm]
100      &0.00  &&0.0043\,(0.0015)   &0.0061\,(0.0016)   &  $-$0.0095\,(0.1463) & $-$0.0027\,(0.1460) &$-$0.0007\,(0.0104) \\
         &0.25  &&0.0030\,(0.0014)   &0.0057\,(0.0014)   &  $-$0.0084\,(0.1547) & $-$0.0184\,(0.1346) &$-$0.0016\,(0.0100)  \\
         &0.50  &&0.0059\,(0.0014)   &0.0022\,(0.0014)   &  $-$0.0205\,(0.1415) & $-$0.0027\,(0.1439) &$-$0.0038\,(0.0059)  \\
         &0.95  &&0.0051\,(0.0015)   &0.0049\,(0.0015)   &  $-$0.0061\,(0.1333) & $-$0.0079\,(0.1364) &$-$0.0004\,(0.0001)  \\[0.1cm]

\noalign{\smallskip}\hline
\end{tabular}
\end{table}

\begin{table}[!ht]
\scriptsize
\centering
\caption{Simulated values of biases and MSEs (within parentheses) of the
ML estimates in comparison with those of MM estimates ($\mu_{k} = 2.0$ and $\delta_{k}=2.0$, for $k=1,2$),
for the BRBS distribution.}\label{tab:04}
\vspace*{0,25cm}
\begin{tabular}{lllrrrrrrrrrr}
\hline\noalign{\smallskip}\\[-0.4cm]
$n$&  &&\multicolumn{5}{c}{ML} \\ \cline{4-8} \\[-0.3cm]
   & \multicolumn{1}{c}{$\rho$}&&\multicolumn{1}{c}{Bias($\widehat{\delta}_{1}$)}&\multicolumn{1}{c}{Bias($\widehat{\delta}_{2}$)}&
                      \multicolumn{1}{c}{Bias($\widehat{\mu}_{1}$)}&\multicolumn{1}{c}{Bias($\widehat{\mu}_{2}$)}&\multicolumn{1}{c}{Bias($\widehat{\rho}$)}\\
\noalign{\smallskip}\hline\noalign{\smallskip}
10       &0.00  &&0.4288\,(4.0119) &0.4007\,(3.8598) & $-$0.0241\,(0.4009) & $-$0.0393\,(0.4058) &$-$0.0037\,(0.1058) \\
         &0.25  &&0.3783\,(3.0608) &0.3574\,(4.2999) & $-$0.0361\,(0.4112) & $-$0.0219\,(0.3867) &$-$0.0356\,(0.1023)  \\
         &0.50  &&0.3771\,(3.4396) &0.3888\,(2.7985) & $-$0.0334\,(0.3855) & $-$0.0621\,(0.3855) &$-$0.0222\,(0.0678)  \\
         &0.95  &&0.3996\,(4.0946) &0.3819\,(4.2157) & $-$0.0383\,(0.3981) & $-$0.0334\,(0.3869) &$-$0.0064\,(0.0025)  \\[0.1cm]
50       &0.00  &&0.0726\,(0.2149) &0.0605\,(0.2057) & $-$0.0064\,(0.0763) & $-$0.0101\,(0.0767) &$-$0.0030\,(0.0196) \\
         &0.25  &&0.0375\,(0.1969) &0.0732\,(0.2061) & $-$0.0076\,(0.0801) & $-$0.0174\,(0.0822) &$-$0.0074\,(0.0170)  \\
         &0.50  &&0.0755\,(0.2147) &0.0873\,(0.2179) & $-$0.0169\,(0.0803) & $-$0.0020\,(0.0775) &$-$0.0015\,(0.0117)  \\
         &0.95  &&0.0548\,(0.2031) &0.0623\,(0.2043) & $-$0.0021\,(0.0782) & $-$0.0026\,(0.0797) &$-$0.0007\,(0.0002)  \\[0.1cm]
100      &0.00  &&0.0316\,(0.0847) &0.0382\,(0.0940) & $-$0.0131\,(0.0369) & $-$0.0065\,(0.0383) &$-$0.0069\,(0.0106) \\
         &0.25  &&0.0386\,(0.0927) &0.0378\,(0.0991) & $-$0.0017\,(0.0429) & $-$0.0015\,(0.0425) &$-$0.0026\,(0.0090)  \\
         &0.50  &&0.0441\,(0.0959) &0.0356\,(0.0955) & $-$0.0036\,(0.0376) & $-$0.0066\,(0.0375) &$-$0.0062\,(0.0061)  \\
         &0.95  &&0.0293\,(0.0837) &0.0344\,(0.0824) & $-$0.0014\,(0.0372) & $-$0.0028\,(0.0366) &$-$0.0003\,(0.0001)  \\[0.1cm]

\hline\noalign{\smallskip}
$n$&  &&\multicolumn{5}{c}{MM} \\ \cline{4-8} \\[-0.3cm]
   & \multicolumn{1}{c}{$\rho$}&&\multicolumn{1}{c}{Bias($\widetilde{\delta}_{1}$)}&\multicolumn{1}{c}{Bias($\widetilde{\delta}_{2}$)}&
                      \multicolumn{1}{c}{Bias($\widetilde{\mu}_{1}$)}&\multicolumn{1}{c}{Bias($\widetilde{\mu}_{2}$)}&\multicolumn{1}{c}{Bias($\widetilde{\rho}$)}\\
\noalign{\smallskip}\hline\noalign{\smallskip}
10       &0.00  &&0.4310\,(4.0159) &0.4030\,(3.8628) & $-$0.0244\,(0.3965) & $-$0.0387\,(0.4119) &$-$0.0038\,(0.1046) \\
         &0.25  &&0.3809\,(3.0624) &0.3598\,(4.3061) & $-$0.0371\,(0.4085) & $-$0.0219\,(0.3885) &$-$0.0368\,(0.1015)  \\
         &0.50  &&0.3819\,(3.4418) &0.3935\,(2.8012) & $-$0.0368\,(0.3752) & $-$0.0653\,(0.3854) &$-$0.0242\,(0.0674)  \\
         &0.95  &&0.4124\,(4.0979) &0.3946\,(4.2191) & $-$0.0382\,(0.4099) & $-$0.0319\,(0.4012) &$-$0.0070\,(0.0025)  \\[0.1cm]
50       &0.00  &&0.0728\,(0.2149) &0.0607\,(0.2057) & $-$0.0055\,(0.0768) & $-$0.0003\,(0.0763) &$-$0.0030\,(0.0195) \\
         &0.25  &&0.0379\,(0.1969) &0.0736\,(0.2061) & $-$0.0070\,(0.0796) & $-$0.0175\,(0.0823) &$-$0.0070\,(0.0169)  \\
         &0.50  &&0.0764\,(0.2147) &0.0882\,(0.2180) & $-$0.0172\,(0.0814) & $-$0.0018\,(0.0783) &$-$0.0021\,(0.0117)  \\
         &0.95  &&0.0573\,(0.2032) &0.0648\,(0.2044) & $-$0.0001\,(0.0797) & $-$0.0048\,(0.0818) &$-$0.0008\,(0.0002)  \\[0.1cm]
100      &0.00  &&0.0317\,(0.0847) &0.0382\,(0.0940) & $-$0.0129\,(0.0370) & $-$0.0065\,(0.0385) &$-$0.0069\,(0.0105) \\
         &0.25  &&0.0387\,(0.0927) &0.0380\,(0.0991) & $-$0.0026\,(0.0430) & $-$0.0024\,(0.0425) &$-$0.0028\,(0.0090)  \\
         &0.50  &&0.0445\,(0.0960) &0.0360\,(0.0955) & $-$0.0027\,(0.0380) & $-$0.0056\,(0.0377) &$-$0.0066\,(0.0061)  \\
         &0.95  &&0.0305\,(0.0838) &0.0356\,(0.0824) & $-$0.0001\,(0.0380) & $-$0.0030\,(0.0376) &$-$0.0003\,(0.0001)  \\[0.1cm]

\noalign{\smallskip}\hline
\end{tabular}
\end{table}

By using the asymptotic distributions given earlier, we obtain the $90\%$ and $95\%$
probability coverages of confidence intervals for the BRBS model.
The $100(1-\gamma)\%$ confidence intervals for $\theta_{l}$, $l=1,\ldots,5$, based
on the ML estimates can be obtained from
 $$
 \left[\left(\widehat{\theta}_{j}+{\frac{z_{\gamma/2}}{\sqrt{{\bm J}_{ll}(\widehat{\bm\theta})}}}\right),
     \left(\widehat{\theta}_{k}+{\frac{z_{1-\gamma/2}}{\sqrt{{\bm J}_{ll}(\widehat{\bm\theta})}}}\right)\right], $$
respectively, where $\widehat{\bm\theta}=(\widehat{\theta}_{1},\widehat{\theta}_{2},\widehat{\theta}_{3},\widehat{\theta}_{4},\widehat{\theta}_{5})^{\top}
=(\widehat{\mu}_{1},\widehat{\mu}_{2},\widehat{\delta}_{1},\widehat{\delta}_{2},\widehat{\rho})^{\top}$
and $z_{r}$ is the $100r$th percentile
of the standard normal distribution.
The $100(1-\gamma)\%$
confidence intervals for $\mu_{k}$ and $\delta_{k}$, $k=1,2$, based on
the MM estimates are given by
\[
\left[\widetilde{\mu}_{k}\left(1+z_{\gamma/2}\sqrt{\frac{h(\widetilde{\delta_{k}})}{n}}\right)^{-1}, \
\widetilde{\mu}_{k}\left(1+z_{1-\gamma/2}\sqrt{\frac{h(\widetilde{\delta_{k}})}{n}}\right)^{-1}\right],\\
\]
\[
\left[\widetilde{\delta}_{k}\left(1+z_{\gamma/2}\sqrt{\frac{2}{n}}\right)^{-1}, \
\widetilde{\delta}_{k}\left(1+z_{1-\gamma/2}\sqrt{\frac{2}{n}}\right)^{-1}\right],
\]
where $h(x)=(2x+5)/{(x+1)^2}$. We compute the $100(1-\gamma)\%$
confidence interval for $\rho$ based on the
MM estimate ($\widetilde{\rho}_{k}$) using both the Fisher's (FI) $z$-transformation
\citep{f:21} and the Krishnamoorthy and Xia's (KX) method \citep{kx:07}; see \cite{kj:15}
for more details about this method. We observe that
\[
X_{k}
=
\sqrt{\frac{\delta_{k}}{2}}\left(\sqrt{\frac{(\delta_{k}+1)T_{k}}{\mu_{k}\delta_{k}}}
-
\sqrt{\frac{\mu_{k}\delta_{k}}{(\delta_{k}+1)T_{k}}} \right) \sim \textrm{N}(0,1), \quad
k=1,2,
\]
and that $\widetilde{\rho}$ as be expressed as
\begin{equation*}\label{appe:C.1}
\widetilde{\rho}
=
\frac{\sum_{i=1}^{n}\prod_{k=1}^{2}x_{ki}}
{\prod_{k=1}^{2}\sqrt{\sum_{i=1}^{n}x_{ki}^2}},
\end{equation*}
where
\[
x_{ki}
=
\sqrt{\frac{\delta_{k}}{2}}
\left(\sqrt{\frac{(\delta_{k}+1)t_{ki}}{\mu_{k}\delta_{k}}}
-
\sqrt{\frac{\mu_{k}\delta_{k}}{(\delta_{k}+1)t_{ki}}} \right),
\quad k=1,2.
\]
It follows that
the pairs $(x_{1i},x_{2i})$, for $i=1,\ldots,n$, can be considered as realizations
of $(X_{1},X_{2})$. Therefore, $\widetilde{\rho}$
is an estimator of the correlation coefficient of a standard bivariate normal distribution.
On the one hand, based on the FI method,
an approximate  $100(1-\gamma)\%$ confidence interval for ${\rho}$ is given by
$$
 \left[\tanh\left(\widetilde{\rho}+\frac{z_{\gamma/2}}{\sqrt{n-3}}\right),\tanh\left(\widetilde{\rho}+\frac{z_{1-\gamma/2}}{\sqrt{n-3}} \right)\right].
$$
On the other hand, based on the KX method, an approximate $100(1-\gamma)\%$ confidence
interval for ${\rho}$ can be obtained from the following steps: 1) obtain $\overline{\rho}={\widetilde{\rho}}/{\sqrt{1-\widetilde{\rho}^2}}$ for a given $n$ and $\widetilde{\rho}$; 2), for $i=1, \ldots,m$ ($m=2,000,000$ say), generate $U_{1}\sim\chi_{n-1}^{2}$, $U_{2}\sim\chi_{n-2}^{2}$ and $Z_{0}\sim \textrm{N}(0,1)$ and compute
\[
Q_{i}
=
\frac{\overline{\rho}\sqrt{U_{2}}-Z_{0}}{\sqrt{(\overline{\rho}\sqrt{U_{2}}-Z_{0})^{2}+U_{1}}}.
\]
Then, the $100(\gamma)$th and $100(1-\gamma)$th
percentiles of the $Q_{i}$'s are the upper and lower limits of $\rho$.

The $90\%$ and $95\%$ probability coverages of confidence intervals are reported in Table~\ref{tab:08}.
From this table, we observe that, the asymptotic confidence intervals
do not present good results for $\delta_{k}$ and $\mu_{k}$ when $n=10$. However, they
improve when $n=50$ and 100. In general,
the coverages for $\rho$ associated with the MM estimates provide better
results compared to the coverages
based on the ML estimates.

\begin{table}[!ht]
\scriptsize
\renewcommand{\arraystretch}{0.9}
\renewcommand{\tabcolsep}{0.1cm}
\centering
\caption{\small Probability coverages of $90\%$ and $95\%$ confidence
intervals for the BRBS model ($\mu_{k}=1.0$, $\delta_{k} = 0.5$, for $k=1,2$).}\label{tab:08}
\vspace*{0,25cm}
\begin{tabular}{lllrrrrrrrrrrrrrrrrrrrrrr}
\hline\noalign{\smallskip}
   &  &&\multicolumn{12}{c}{ML} \\ \cline{4-16}
$n$&  &&\multicolumn{6}{c}{$90\%$}  &&\multicolumn{6}{c}{$95\%$} \\ \cline{4-9} \cline{11-16}
   & \multicolumn{1}{c}{$\rho$}&&\multicolumn{1}{c}{${\delta}_{1}$}&\multicolumn{1}{c}{${\delta}_{2}$}&
                      \multicolumn{1}{c}{${\mu}_{1}$}&\multicolumn{1}{c}{${\mu}_{2}$}&\multicolumn{2}{c}{${\rho}$}
                      &&\multicolumn{1}{c}{${\delta}_{1}$}&\multicolumn{1}{c}{${\delta}_{2}$}&
                      \multicolumn{1}{c}{${\mu}_{1}$}&\multicolumn{1}{c}{${\mu}_{2}$}&\multicolumn{2}{c}{${\rho}$}
                      \\
\noalign{\smallskip}\hline\noalign{\smallskip}
10       &0.00  &&  78.90 & 79.32 & 82.74 & 81.76 & \multicolumn{2}{c}{76.98} && 84.24 & 84.18 & 88.00 & 87.96 & \multicolumn{2}{c}{84.18}\\
         &0.25  &&  79.52 & 78.44 & 81.08 & 82.12 & \multicolumn{2}{c}{77.44} && 84.34 & 83.96 & 86.28 & 87.66 & \multicolumn{2}{c}{83.16}\\
         &0.50  &&  78.88 & 79.68 & 75.58 & 75.44 & \multicolumn{2}{c}{77.42} && 84.08 & 83.62 & 83.30 & 84.06 & \multicolumn{2}{c}{83.98}\\
         &0.95  &&  78.64 & 78.82 & 37.58 & 36.26 & \multicolumn{2}{c}{79.24} && 83.63 & 82.68 & 41.75 & 44.40 & \multicolumn{2}{c}{83.42}\\[0.1cm]
50       &0.00  &&  87.66 & 88.58 & 87.64 & 89.30 & \multicolumn{2}{c}{88.56} && 92.24 & 92.62 & 93.32 & 93.06 &  \multicolumn{2}{c}{93.20} \\
         &0.25  &&  87.16 & 88.18 & 87.14 & 86.56 & \multicolumn{2}{c}{87.28} && 92.44 & 92.78 & 93.06 & 92.18 &  \multicolumn{2}{c}{93.54}  \\
         &0.50  &&  87.34 & 87.86 & 83.02 & 83.08 & \multicolumn{2}{c}{88.12} && 93.38 & 93.18 & 90.42 & 90.14 &  \multicolumn{2}{c}{93.54}  \\
         &0.95  &&  88.54 & 88.26 & 38.54 & 37.82 & \multicolumn{2}{c}{88.28} && 93.32 & 92.86 & 44.20 & 44.74 &  \multicolumn{2}{c}{93.32}  \\[0.1cm]
100      &0.00  &&  89.00 & 89.70 & 89.30 & 89.10 & \multicolumn{2}{c}{87.50} && 94.00 & 95.10 & 94.10 & 94.40 &  \multicolumn{2}{c}{92.90} \\
         &0.25  &&  89.10 & 88.90 & 89.20 & 88.60 & \multicolumn{2}{c}{90.00} && 95.10 & 92.30 & 93.60 & 92.10 &  \multicolumn{2}{c}{93.80}  \\
         &0.50  &&  89.30 & 89.80 & 83.90 & 84.40 & \multicolumn{2}{c}{90.50} && 92.60 & 94.10 & 90.50 & 89.80 &  \multicolumn{2}{c}{93.10}  \\
         &0.95  &&  90.10 & 89.90 & 38.44 & 38.44 & \multicolumn{2}{c}{88.20} && 93.00 & 94.30 & 48.00 & 48.50 &  \multicolumn{2}{c}{93.80}  \\[0.1cm]

\hline\noalign{\smallskip}
   &  &&\multicolumn{12}{c}{MM} \\ \cline{4-16}
$n$&  &&\multicolumn{6}{c}{$90\%$}  &&\multicolumn{6}{c}{$95\%$} \\ \cline{4-9} \cline{11-16}
   & \multicolumn{1}{c}{$\rho$}&&\multicolumn{1}{c}{${\delta}_{1}$}&\multicolumn{1}{c}{${\delta}_{2}$}&
                      \multicolumn{1}{c}{${\mu}_{1}$}&\multicolumn{1}{c}{${\mu}_{2}$}&\multicolumn{1}{c}{${\rho}$ (FI)}&\multicolumn{1}{c}{${\rho}$ (KX)}
                      &&\multicolumn{1}{c}{${\delta}_{1}$}&\multicolumn{1}{c}{${\delta}_{2}$}&
                      \multicolumn{1}{c}{${\mu}_{1}$}&\multicolumn{1}{c}{${\mu}_{2}$}&\multicolumn{1}{c}{${\rho}$ (FI)}&\multicolumn{1}{c}{${\rho}$ (KX)}
                      \\
\noalign{\smallskip}\hline\noalign{\smallskip}
10       &0.00  && 78.90  & 79.34 & 84.94 & 83.96 & 90.42 & 89.46 && 84.22 & 84.16 & 89.82 & 89.52 & 94.54 & 95.78\\
         &0.25  && 79.50  & 78.48 & 84.64 & 84.88 & 90.08 & 90.28 && 84.32 & 83.96 & 89.28 & 90.24 & 95.48 & 95.14\\
         &0.50  && 78.78  & 79.62 & 83.98 & 84.06 & 90.14 & 89.44 && 84.04 & 83.58 & 89.98 & 90.00 & 94.94 & 95.18\\
         &0.95  && 78.58  & 78.86 & 84.64 & 83.94 & 90.64 & 90.28 && 83.53 & 82.68 & 90.81 & 89.65 & 95.62 & 94.61\\[0.1cm]
50       &0.00  && 87.66  & 88.58 & 88.14 & 89.52 & 90.56 & 90.60 && 92.24 & 92.62 & 93.58 & 93.46 & 95.02 & 94.82\\
         &0.25  && 87.16  & 88.16 & 88.78 & 88.40 & 89.96 & 89.68 && 92.44 & 92.74 & 94.02 & 93.34 & 95.16 & 95.42\\
         &0.50  && 87.32  & 87.86 & 89.04 & 89.38 & 90.08 & 90.10 && 93.40 & 93.18 & 94.32 & 94.46 & 95.32 & 95.72\\
         &0.95  && 88.48  & 88.24 & 88.24 & 88.12 & 90.50 & 90.46 && 93.24 & 92.92 & 93.64 & 93.64 & 95.36 & 95.14\\[0.1cm]
100      &0.00  && 89.00  & 89.70 & 90.00 & 89.70 & 90.30 & 88.30 && 94.03 & 95.10 & 94.20 & 94.40 & 94.60 & 94.20\\
         &0.25  && 89.10  & 88.90 & 90.70 & 89.50 & 89.30 & 89.90 && 95.12 & 92.30 & 94.70 & 93.70 & 95.60 & 94.90\\
         &0.50  && 89.40  & 89.80 & 89.50 & 90.40 & 88.30 & 90.90 && 92.65 & 94.19 & 95.10 & 94.60 & 96.50 & 94.20\\
         &0.95  && 90.00  & 89.80 & 89.90 & 89.30 & 90.10 & 89.80 && 93.00 & 94.38 & 95.00 & 95.00 & 95.10 & 95.30\\[0.1cm]
\noalign{\smallskip}\hline
\end{tabular}
\end{table}

\subsection{Real-world reliability data analysis}\label{sec:6}

\begin{figure}[h!]
\centering
\psfrag{0.0}[c][c]{\scriptsize{0.0}}
\psfrag{0.1}[c][c]{\scriptsize{0.1}}
\psfrag{0.2}[c][c]{\scriptsize{0.2}}
\psfrag{0.3}[c][c]{\scriptsize{0.3}}
\psfrag{0.4}[c][c]{\scriptsize{0.4}}
\psfrag{0.5}[c][c]{\scriptsize{0.5}}
\psfrag{0.6}[c][c]{\scriptsize{0.6}}
\psfrag{0.7}[c][c]{\scriptsize{0.7}}
\psfrag{0.8}[c][c]{\scriptsize{0.8}}
\psfrag{1.0}[c][c]{\scriptsize{1.0}}
\psfrag{0.000}[c][c]{\scriptsize{0.0}}
\psfrag{0.002}[c][c]{\scriptsize{.002}}
\psfrag{0.004}[c][c]{\scriptsize{.004}}
\psfrag{0.006}[c][c]{\scriptsize{.006}}
\psfrag{0.008}[c][c]{\scriptsize{.008}}
\psfrag{0.010}[c][c]{\scriptsize{.010}}
\psfrag{0.00}[c][c]{\scriptsize{0.00}}
\psfrag{0.05}[c][c]{\scriptsize{0.05}}
\psfrag{0.10}[c][c]{\scriptsize{0.10}}
\psfrag{0.15}[c][c]{\scriptsize{0.15}}
\psfrag{0.20}[c][c]{\scriptsize{0.20}}
\psfrag{0}[c][c]{\scriptsize{0}}
\psfrag{2}[c][c]{\scriptsize{2}}
\psfrag{4}[c][c]{\scriptsize{4}}
\psfrag{6}[c][c]{\scriptsize{6}}
\psfrag{8}[c][c]{\scriptsize{8}}
\psfrag{10}[c][c]{\scriptsize{10}}
\psfrag{100}[c][c]{\scriptsize{100}}
\psfrag{200}[c][c]{\scriptsize{200}}
\psfrag{300}[c][c]{\scriptsize{300}}
\psfrag{400}[c][c]{\scriptsize{400}}
\psfrag{bs}[c][c]{\scriptsize{BS CDF}}
\psfrag{bst}[c][c]{\scriptsize{BS-$t$ CDF}}
\psfrag{em}[c][c]{\scriptsize{Empirical CDF}}
\psfrag{w}[c][c]{\scriptsize{$W_{n}(k/n)$}}
\psfrag{n}[c][c]{\scriptsize{$k/n$}}
\subfigure[RBS PP - shock ($t_{1}$)]    {\includegraphics[height=3.8cm,width=3.8cm,angle=-90]{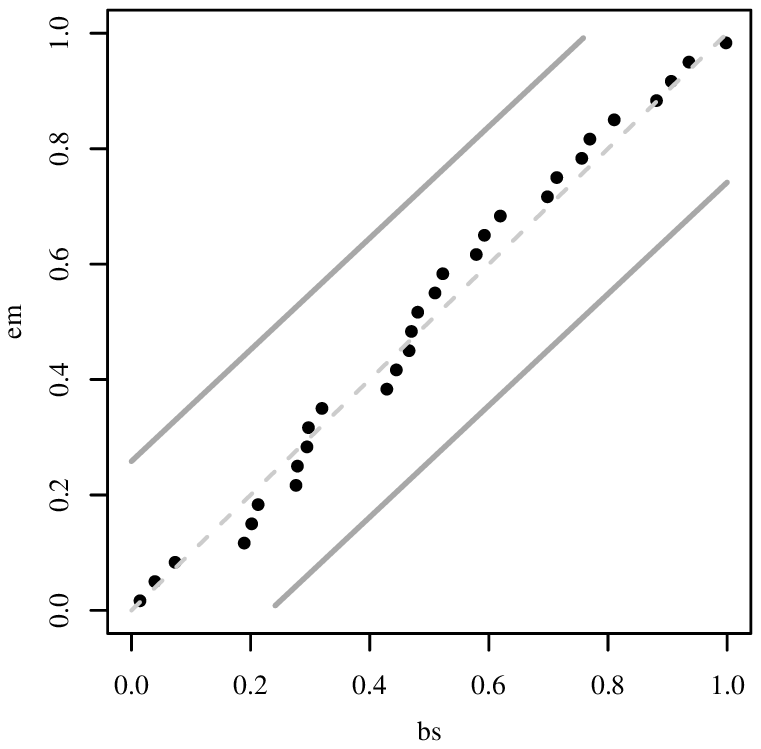}}
\subfigure[RBS PP - vibration  ($t_{2}$)]     {\includegraphics[height=3.8cm,width=3.8cm,angle=-90]{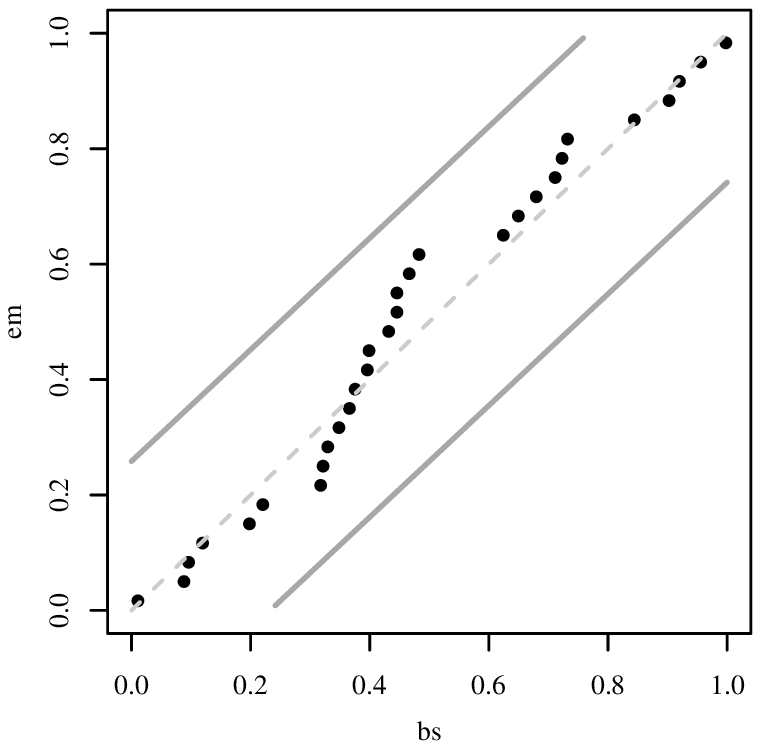}}
\subfigure[TTT - shock  ($t_{1}$)]      {\includegraphics[height=3.8cm,width=3.8cm,angle=-90]{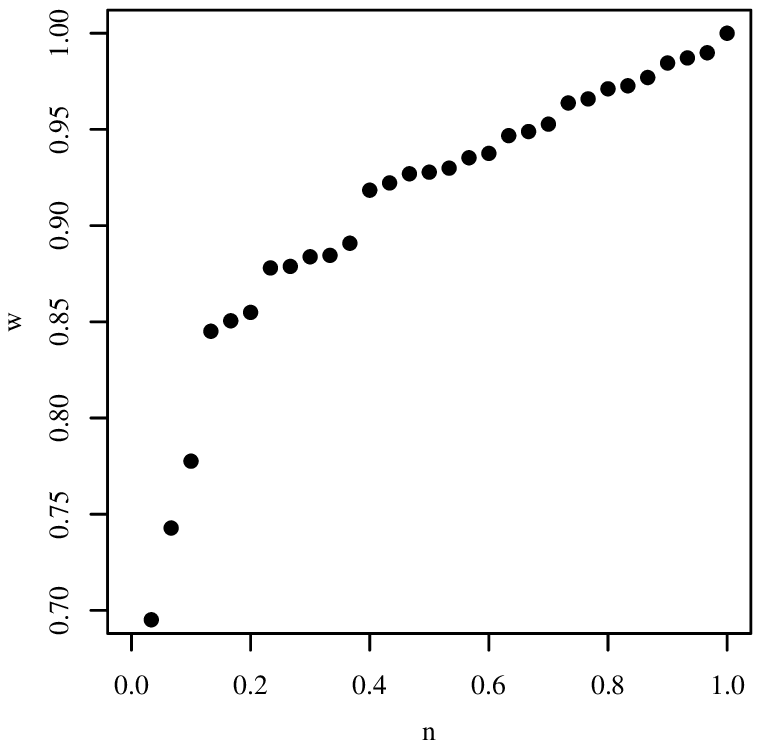}}
\subfigure[TTT - vibration  ($t_{2}$)]       {\includegraphics[height=3.8cm,width=3.8cm,angle=-90]{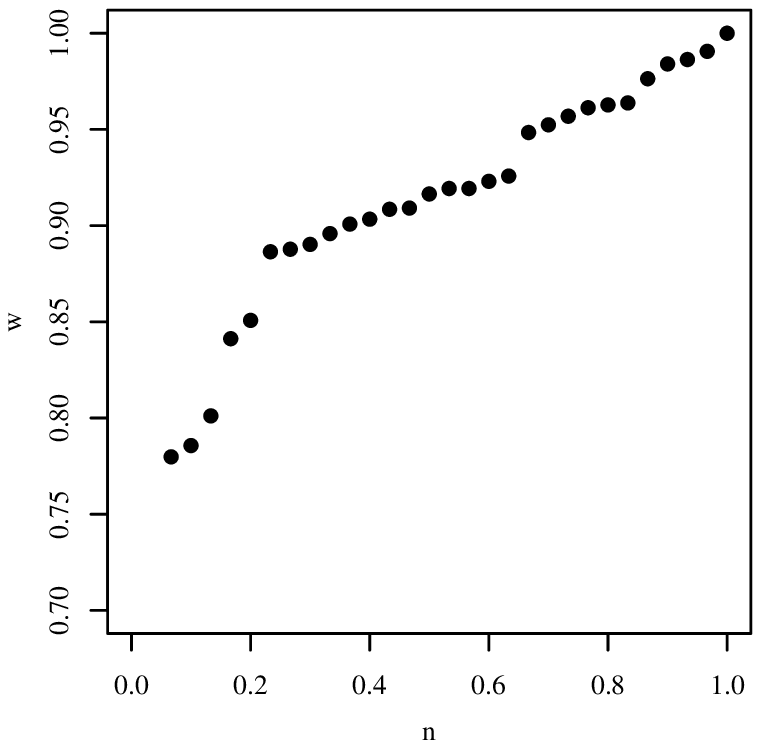}}
\caption{\small PP plots with acceptance bands and scaled TTT plots with the stiffness data.}
  \label{fig:1}
\end{figure}

We here illustrate the BRBS distribution by using a real data set, which corresponds to two different measurements of stiffness, namely,
shock and vibration each of $n=30$ boards. The former refers to the emission of shock wave down the board, while the
latter is obtained during the vibration of the board; see \cite{jw:07}. We
consider probability versus probability (PP) plots with acceptance
bands based on the marginal distributions of $T_{1}$ and $T_{2}$ to support
the BRBS model; see Figure~\ref{fig:1}(a)-(b).
We also consider the scaled total time on test (TTT) plots in
Figure~\ref{fig:1}(c)-(d) to have
an idea about the shape of the HR of the marginals;
see \cite{alab:12}.
From Figure~\ref{fig:1}, we observe that the PP plots support the
BRBS model and the TTT plots suggest that both marginals
have unimodal failure rates.

\begin{figure}[h!]
\centering
\vspace{-0.3cm}
\psfrag{0.0}[c][c]{\scriptsize{0.0}}
\psfrag{0.1}[c][c]{\scriptsize{0.1}}
\psfrag{0.2}[c][c]{\scriptsize{0.2}}
\psfrag{0.3}[c][c]{\scriptsize{0.3}}
\psfrag{0.4}[c][c]{\scriptsize{0.4}}
\psfrag{0.5}[c][c]{\scriptsize{0.5}}
\psfrag{0.6}[c][c]{\scriptsize{0.6}}
\psfrag{0.7}[c][c]{\scriptsize{0.7}}
\psfrag{0.8}[c][c]{\scriptsize{0.8}}
\psfrag{1.0}[c][c]{\scriptsize{1.0}}
\psfrag{0.000}[c][c]{\scriptsize{0.0}}
\psfrag{0.002}[c][c]{\scriptsize{.002}}
\psfrag{0.004}[c][c]{\scriptsize{.004}}
\psfrag{0.006}[c][c]{\scriptsize{.006}}
\psfrag{0.008}[c][c]{\scriptsize{.008}}
\psfrag{0.010}[c][c]{\scriptsize{.010}}
\psfrag{0.00}[c][c]{\scriptsize{0.00}}
\psfrag{0.05}[c][c]{\scriptsize{0.05}}
\psfrag{0.10}[c][c]{\scriptsize{0.10}}
\psfrag{0.15}[c][c]{\scriptsize{0.15}}
\psfrag{0.20}[c][c]{\scriptsize{0.20}}
\psfrag{0}[c][c]{\scriptsize{0}}
\psfrag{2}[c][c]{\scriptsize{2}}
\psfrag{4}[c][c]{\scriptsize{4}}
\psfrag{6}[c][c]{\scriptsize{6}}
\psfrag{8}[c][c]{\scriptsize{8}}
\psfrag{10}[c][c]{\scriptsize{10}}
\psfrag{100}[c][c]{\scriptsize{100}}
\psfrag{200}[c][c]{\scriptsize{200}}
\psfrag{300}[c][c]{\scriptsize{300}}
\psfrag{400}[c][c]{\scriptsize{400}}
\psfrag{0.4435}[c][c]{\tiny{0.4435}}
\psfrag{EC}[c][c]{\scriptsize{Empirical Quantile}}
\psfrag{TC}[c][c]{\scriptsize{Theoretical Quantile}}
\psfrag{ks}[r][r]{\tiny{KS p-value}}
{\includegraphics[height=5.2cm,width=5.2cm,angle=-90]{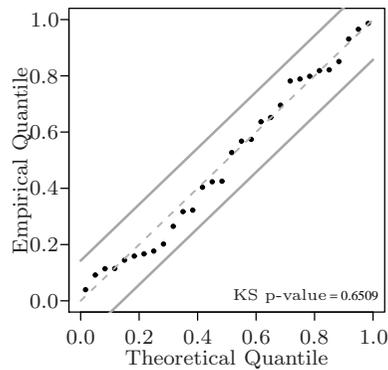}}
\vspace{-0.3cm}
\caption{\small PP plots with acceptance bands at 5\% for the transformed Mahalanobis distance with the stiffness data.}
  \label{fig:MahaPPa}
\end{figure}

We now fit the BRBS distribution to the stiffness data
set. From the observations, we obtain
$s_{1}=1906.10$, $r_{1}=1857.55$, $s_{2}=1749.53$ and $r_{2}=1699.99$. Table~\ref{tab:10}
presents the ML and MM estimates along
with their corresponding SEs and 95\% confidence intervals, as well as the
log-likelihood values. We note that across the models both the estimates
and log-likelihood values are quite similar. The Mahalanobis distance can
be used to check the validity of the
BRBS model; see \cite{vbz:14b}. In the BRBS case, this distance is given by
\begin{equation}\label{mahadis}
 D_{i}({\bm\theta})={\bm\xi}_{i}^{\top}{\bm\psi}^{-1}{\bm\xi}_{i},
\end{equation}
for $i=1,\ldots,n$, where
$
{\bm\psi}
=
\begin{pmatrix}
1 & \rho \\
\rho & 1
\end{pmatrix},
$
$
{\bm\xi}_{i}
=
\big((a_{1i}-b_{1i})\sqrt{{\delta_{1}}/{2}},
(a_{2i}-b_{2i}) \sqrt{{\delta_{2}}/{2}}\big)^{\top}
$
with $a_{1i}$, $b_{1i}$, $a_{2i}$ and
$b_{2i}$ defined as in \eqref{sec2.1:01}. Based on \cite{mlc:15}, it
follows that the Mahalanobis distance, with ${\bm\theta}$ substituted by its ML estimator
$\widehat{\bm\theta}$, has asymptotically a $\chi^{2}_{2}$ distribution. We apply the Wilson-Hilferty approximation for
transforming to normality the Mahalanobis distance defined in \eqref{mahadis}. Then,
goodness of fit of the BRBS model can be assessed by checking
normality of the transformed distances with the Wilson-Hilferty approximation;
see \cite{ipg:14}. Figure \ref{fig:MahaPPa} shows the PP plot
with acceptance bands of the transformed Mahalanobis distance for the BRBS
distribution. From this figure, observe that the considered model
provides a good fit, which is confirmed by the associated $p$-value 0.4433
of the Kolmogorov-Smirnov (KS) test.

\begin{table}[h!]
 \scriptsize
\centering
\caption{ML and MM estimates of the BRBS model parameters, SEs and the corresponding
95\% confidence intervals with the stiffness data.}\label{tab:10}
\vspace*{0.25cm}
\begin{tabular}{lcccccccccccccc}
\hline
Parameter         &  &&      ML       &  SE      &   95\% Conf. Interval        &&      MM     &  SE   &   95\% Conf. Interval             \\ \hline
$\mu_{1}$         &  &&   1906.100    & 56.826   &(1885.766,1926.435)&&   1906.100  &56.247 &(1795.857,2016.343)           \\
$\mu_{2}$         &  &&   1749.533    & 55.092   &(1729.819,1769.247)&&   1749.533  &54.513 &(1642.688,1856.378)            \\
$\delta_{1}$      &  &&   77.030      & 19.967   & (69.885,76.279)   &&   77.030    & 19.889 &(38.048,108.116)              \\
$\delta_{2}$      &  &&   69.134      & 17.924   & (62.720,75.548)   &&   69.134    & 17.850 &  (34.148,104.121)             \\
$\rho$            &  &&    0.908      & 0.032    &  (0.897,0.920)    &&    0.908    & -- &  (0.814,0.956)                \\[0.15cm]
 $\ell({\bm\theta})$&&&   $-$400.648  &     &                        && $-$400.648  &       &                               \\
\hline
\end{tabular}
\end{table}

\section{Concluding remarks}

In this paper we have proposed a new bivariate Birnbaum-Saunders distribution, which is established in terms of its means. We have discussed
several properties of the reparameterized bivariate Birnbaum-Saunders model including unimodality of the probability density function and monotonicity of the hazard rate. Moreover, we have
discussed maximum likelihood estimation and modified moment estimation of the model parameters. Numerical results have illustrated the potentiality of the proposed model. As part of future work, it would be of interest to develop likelihood
inferential methods by considering censored data. Moreover, implementation in a regression context would be of practical relevance. Finally, time
series models based on the proposed bivariate distribution with corresponding influence diagnostic tools can
also be considered; see \cite{saulolla:18}. Work on these problems is currently
in progress and we hope to report these findings in a future paper.

\section{Appendix: proofs}

\small
\begin{proof}[Lemma \ref{pre-lemma}]
1. The proof is immediate since $a_k(\cdot)$, $k=1,2$, is a strictly increasing function. 2. By hypothesis $a_1(t)\geqslant \sqrt{1-\rho^2} \,(1+\rho)$ and $a_2(w)>\sqrt{1-\rho^2}$. Since $\rho<0$ we have $-\rho a_2(w)>-\rho\sqrt{1-\rho^2}$.
Therefore, $a_1(t)-\rho a_2(w)\geqslant \sqrt{1-\rho^2}$. 3. By hypothesis $a_1(t)\geqslant \sqrt{1-\rho^2}.$ Since $w<\beta_2$ $(>\beta_2)$ and
$\rho>0$ $(<0)$ we have $-\rho a_2(w)>0$. Therefore, $a_1(t)-\rho a_2(w)\geqslant \sqrt{1-\rho^2}.$ Finally, the proof of Item 4 (Item 5) follows by combining Items 1 and 2 (Items 1 and 3).
\end{proof}

\begin{proof}[Proposition \ref{prop-del}]
Using the expression \eqref{sec2:02} for the PDF of the \textrm{BRBS}
distribution, we obtain
\begin{eqnarray}
\frac{\partial}{\partial t_1} f_{T_1,T_2}(t_1,t_2)
&=&
f_{T_1,T_2}(t_1,t_2)[a_1'(t_1)]^2
\left\{
\frac{a_1''(t_1)}{[a_1'(t_1)]^2}
-
\frac{c_{1,2}(t_1,t_2;\rho)}{\sqrt{1-\rho^2}}
\right\},\label{der-1}\\
\frac{\partial}{\partial t_2} f_{T_1,T_2}(t_1,t_2)
&=&
f_{T_1,T_2}(t_1,t_2)[a_2'(t_2)]^2
\left\{
\frac{a_2''(t_2)}{[a_2'(t_2)]^2}
-
\frac{c_{2,1}(t_2,t_1;\rho)}{\sqrt{1-\rho^2}}
\right\},   \label{der-2}
\end{eqnarray}
where $c_{j,k}(s,t;\rho)$ is defined in \eqref{cp}. Then,
\begin{align}\label{grad}
\nabla f_{T_1,T_2}(t_1,t_2) = 0
\quad \textrm{iff} \quad
\frac{a_1''(t_1)}{[a_1'(t_1)]^2}
-
\frac{c_{1,2}(t_1,t_2;\rho)}{\sqrt{1-\rho^2}}
\ = \
\frac{a_2''(t_2)}{[a_2'(t_2)]^2}
-
\frac{c_{2,1}(t_2,t_1;\rho)}{\sqrt{1-\rho^2}}
\ = \ 0.
\end{align}
In what follows we will prove that if the point
$(\widetilde{t}_1,\widetilde{t}_2)=(c\beta_1,c\beta_2)$ is critical for $f_{T_1,T_2}$,
then the constant $c>0$ is unique. In fact, since $(\widetilde{t}_1,\widetilde{t}_2)$
is a critical point, by \eqref{grad} and by expressions of the first and second derivatives
of $a(\cdot)$ (provided in \eqref{rem-us}) we have $p(c)=0$ e $q(c)=0$,
where $p$ and $q$ are the cubic polynomials given by
\begin{align*}
p(c)=
c^3+\left(1+\frac{\alpha_1}{\theta_{1,2}}\right)c^2
+\left(\frac{3\alpha_1}{\theta_{1,2}}-1\right)c-1,
\quad
q(c)=
c^3+\left(1+\frac{\alpha_2}{\theta_{2,1}}\right)c^2
+\left(\frac{3\alpha_2}{\theta_{2,1}}-1\right)c-1
\end{align*}
and $\theta_{j,k}=(1/(1-\rho^2))({1/\alpha_j}-\rho/\alpha_k)$.
By condition 1 of the Hypothesis 1, $\theta_{j,k}>0$.

From here on we analyze only the polynomial $p(c)$, since from the analysis of this polynomial we obtain the same conclusion.

The discriminant of a cubic polynomial $ax^3+bx^2+cx+d$ is given by $\Delta=b^2c^2-4ac^3-4b^3d-27a^2d^2+18abcd$. In our case, the discriminant of $p$ is given by
\[
\begin{array}{c}
\Delta_p
=
\left(1+\frac{\alpha_1}{\theta_{1,2}}\right)^2 \!
\left(\frac{3\alpha_1}{\theta_{1,2}}-1\right)^2
-
4 \left(\frac{3\alpha_1}{\theta_{1,2}}-1\right)^3\!
+
4 \left(1+\frac{\alpha_1}{\theta_{1,2}}\right)^3\!
-
18 \left(1+\frac{\alpha_1}{\theta_{1,2}}\right)
\left(\frac{3\alpha_1}{\theta_{1,2}}-1\right)
-
27.
\end{array}
\]
By condition 2 of the Hypothesis 1, it can be seen that $\Delta_p>0$. Then, the
polynomial $p(c)$ has three distinct real roots, denoted $c_1,c_2$ and $c_3$.
By Vieta's formula, it is valid that
$$
c_1+c_2+c_3=-\left(1+\frac{\alpha_1}{\theta_{1,2}}\right), \quad c_1 \,c_2+c_1 \,c_3+c_2 \,c_3= \frac{3\alpha_1}{\theta_{1,2}}-1, \quad c_1 \,c_2 \,c_3= 1.
$$%
From the first and third equations above, we conclude that there must be
two negative and one positive roots. The proof is complete.
\end{proof}

\begin{proof}[Theorem \ref{theo:1}]
By Proposition \ref{prop-del} we must only prove that $(\widetilde{t}_1,\widetilde{t}_2)$
is a maximum point for $f_{T_1,T_2},$ whenever $c\in(0,2\sqrt{3}-3)$.
For this, deriving in \eqref{der-1} and \eqref{der-2}, for $k=1,2,$ we obtain
\begin{eqnarray*}
\frac{\partial^2}{\partial t_k^2} f_{T_1,T_2}(t_1,t_2)
\Big|_{t_1=\widetilde{t}_1,\, t_2=\widetilde{t}_2}
&=&
f_{T_1,T_2}(\widetilde{t}_1,\widetilde{t}_2)
\left\{
\frac{a_k'''(\widetilde{t}_k) a_k'(\widetilde{t}_k)-2[a_k''(\widetilde{t}_k)]^2 }
{[a_k'(\widetilde{t}_k)]^2}
-
\frac{[a_k'(\widetilde{t}_k)]^2}{1-\rho^2}
\right\},
\\
\frac{\partial^2}{\partial t_2\partial t_1} f_{T_1,T_2}(t_1,t_2)
\Big|_{t_1=\widetilde{t}_1,\, t_2=\widetilde{t}_2}
&=&
\left(\frac{\rho}{1-\rho^2}\right)
f_{T_1,T_2}(\widetilde{t}_1,\widetilde{t}_2)\,
a_1'(\widetilde{t}_1)
a_2'(\widetilde{t}_2).
\end{eqnarray*}
Denoting
\begin{eqnarray*}
\Delta
&=&
\prod_{k=1}^{2}
\frac{\partial^2}{\partial t_k^2} f_{T_1,T_2}(t_1,t_2)
\Big|_{t_1=\widetilde{t}_1,\, t_2=\widetilde{t}_2}
-
\left[
\frac{\partial^2}{\partial t_2\partial t_1} f_{T_1,T_2}(t_1,t_2)
\Big|_{t_1=\widetilde{t}_1,\, t_2=\widetilde{t}_2}
\right]^2
\end{eqnarray*}
and using the above identities we have
\begin{eqnarray*}
\Delta
&=&
\left[f_{T_1,T_2}(\widetilde{t}_1,\widetilde{t}_2)\right]^2
\left\{
[a_1'(\widetilde{t}_1) a_2'(\widetilde{t}_2)]^2
-
\frac{[a_1'(\widetilde{t}_1)]^2}{1-\rho^2} \
\frac{a_2'''(\widetilde{t}_2) a_2'(\widetilde{t}_2)-2[a_2''(\widetilde{t}_2)]^2 }
{[a_2'(\widetilde{t}_2)]^2}
\right.
\\
&&
\left.
-
\frac{[a_2'(\widetilde{t}_2)]^2}{1-\rho^2} \
\frac{a_1'''(\widetilde{t}_1) a_1'(\widetilde{t}_1)-2[a_1''(\widetilde{t}_1)]^2 }
{[a_1'(\widetilde{t}_1)]^2}
+
\prod_{k=1}^{2}
\frac{a_k'''(\widetilde{t}_k) a_k'(\widetilde{t}_k)-2[a_k''(\widetilde{t}_k)]^2 }
{[a_k'(\widetilde{t}_k)]^2}
\right\}.
\end{eqnarray*}
By second derivative criteria for maxima and minima, the point
$(\widetilde{t}_1,\widetilde{t}_2)$ is of maximum for $f_{T_1,T_2}$ whenever
$a_k'''(\widetilde{t}_k) a_k'(\widetilde{t}_k)-2[a_k''(\widetilde{t}_k)]^2<0$, for $k=1,2$.
Using the expressions of the first, second and third derivatives of $a(\cdot)$
(provide in \eqref{rem-us}), a straightforward calculus shows that the inequality
above is true if and only if
$\widetilde{t}^2_k+6\beta_k \widetilde{t}_k-3\beta_k<0$
$\Leftrightarrow$
$c^2+6c-3<0$ $\Leftrightarrow$
$c\in(0,2\sqrt{3}-3)$, completing the proof.
\end{proof}

\begin{proof}[Proposition \ref{Marginal}]
Since $Q(u,v)=v^2+\big[(u-\rho v)/\sqrt{1-\rho^2}\big]^2$, we have
\begin{align}\label{new-density}
f_{T_{1},T_{2}}(t_{1},t_{2})
=
\frac{1}{2\pi\sqrt{1-\rho^2}} \exp\left(-\frac{a_2^2(t_2)}{2}\right) \, a_2'(t_2) \cdot \,
\exp\left(-\frac{1}{2}\{c_{1,2}(t_1,t_2;\rho)\}^2\right)  \, a_1'(t_1),
\end{align}
where $c_{1,2}(t_1,t_2;\rho)$ is defined in \eqref{cp}.
Then,
\begin{align*}
f_{T_2}(t_2)
&=
\frac{1}{2\pi\sqrt{1-\rho^2}} \ \exp\left(-\frac{a_2^2(t_2)}{2}\right) a_2'(t_2)
\int_0^\infty
\exp\left(-\frac{1}{2}\{c_{1,2}(t_1,t_2;\rho)\}^2\right)  \, a_1'(t_1) \,\textrm{d}t_1
\\
&=
\frac{1}{\sqrt{2\pi}} \ \exp\left(-\frac{a_2^2(t_2)}{2}\right) a_2'(t_2).
\end{align*}
Analogously, we prove that $f_{T_1}(t_1)=f(t_1;\mu_1,\delta_1)$.
This completes the proof.
\end{proof}
\begin{proof}[Proposition \ref{Reliability function}]
Using \eqref{new-density}, for $r>0$, we have
\begin{equation}\label{pre-call}
I(r,w)
\ = \
\int_{0}^{r} f_{T_1,T_2}(w,t) \,\textrm{d}t
=
\frac{1}{\sqrt{2\pi}} \ \exp\left(-\frac{a_1^2(w)}{2}\right) a_1'(w) \
\Phi\big(c_{2,1}(r,w;\rho)\big)  =
f(w;\mu_1,\delta_1) \
\Phi\big(c_{2,1}(r,w;\rho)\big).
\end{equation}
Combining the identity $\textrm{P}(T_1<T_2)=\int_{0}^{\infty}I(w,w)\, dw$,
the above relations and the fact that $T_1\sim\textrm{RBS}(\mu_1,\delta_1)$
(by Proposition \ref{Marginal}), we obtain
\begin{align*}
R
=
\int_{0}^{\infty} f(w;\mu_1,\delta_1)
\Phi\big(c_{2,1}(w,w;\rho)\big) \, \textrm{d}w
=
\textrm{E}\left[\Phi\big(c_{2,1}(T_1,T_1;\rho)\big)\right],
\end{align*}
completing the proof.
\end{proof}

\begin{proof}[Proposition \ref{prop-pre}]
To prove the Items 1 and 2, we simply use a similar result from \eqref{pre-call}.
\end{proof}

\begin{proof}[Proposition \ref{prop:pre}]
Let 
$
g(w,t_2)=f(w;\mu_1,\delta_1)\left\{1-\Phi\big(c_{2,1}(t_2,w;\rho)\big) \right\}.
$
Since $w\mapsto g(w,t_2)$ is a nonnegative function, follows that 
$t_1\mapsto G(t_1,t_2)=\int_{t_1}^\infty g(w,t_2)\,\textrm{d}w$ is a decreasing function. 
By other hand, since 
$\Phi(\cdot)$ is a CDF and $t_2\mapsto c_{2,1}(t_2,w;\rho)$ is an increasing function, 
immediately follows that $G(t_1,t_2)$ is also decreasing in  $t_2$.
\end{proof}

\begin{proof}[Proposition \ref{prop:3.5}]
1. Let $L(t_1,t_2)=f(t_1;\mu_1,\delta_1) \phi\big(c_{2,1}(t_2,t_1;\rho)\big)$.	
The hypotheses
$\rho<0$, $t_1\leqslant \beta_1$ and $t_2\leqslant \beta_2$ imply that
$a_1(t_1)\leqslant 0$ and $c_{2,1}(t_2,t_1;\rho)\leqslant 0$. 
Then $t_1\mapsto f(t_1;\mu_1,\delta_1)$ and $t_1\mapsto\phi\big(c_{2,1}(t_2,t_1;\rho)\big)$
are increasing functions. Hence, by Proposition \ref{prop:pre} the function
$t_1\mapsto h_{T_1,T_2}(t_1,t_2)=L(t_1,t_2)/G(t_1,t_2)$ is increasing
since the product of nonnegative increasing functions is also increasing, where
$G(t_1,t_2)$ was defined in Proposition \ref{prop:pre}.
In order to verify Item 2, see that
the conditions 
$\rho> 0$, $t_1\geqslant \beta_1$ and $t_2\leqslant \beta_2$ imply that 
$c_{2,1}(t_2,t_1;\rho)\leqslant 0$. Then
$t_2\mapsto\phi\big(c_{2,1}(t_2,t_1;\rho)\big)$
is an increasing function. So, by Proposition \ref{prop:pre}  the function
$t_2\mapsto h_{T_1,T_2}(t_1,t_2)=L(t_1,t_2)/G(t_1,t_2)$ is increasing.
Finally, to prove Item 3, note that the assumptions
$\rho= 0$ and $t_2\leqslant \beta_2$ imply that 
$c_{2,1}(t_2,t_1;\rho)=a_2(t_2)\leqslant 0$. 
Then
$t_2\mapsto\phi\big(c_{2,1}(t_2,t_1;\rho)\big)$ is an increasing function 
and the proof follows in analogy to Item 2.
\end{proof}

\begin{proof}[Proposition \ref{prop:3.6}]
Using \eqref{exp-hr} we obtain
\begin{align*}
h_{T_1,T_2}(t_1,t_2)
\geqslant
\frac{f(t_1;\mu_1,\delta_1) \phi\big(c_{2,1}(t_2,t_1;\rho)\big)}
{1-\textrm{E}\left[\Phi\big(c_{2,1}(t_2,T_1;\rho)\big) \right]}.
\end{align*}
Taking $t_2=w$ on the above equality and using the Proposition \ref{Reliability function},
we get
\[
h_{T_1,T_2}(t_1,t_2=w)
\geqslant
\frac{f(t_1;\mu_1,\delta_1) \phi\big(c_{2,1}(t_2,t_1;\rho)\big)}
{1-R}.
\]
Finally, taking $w=t_1$, integrating from $0$ to $\infty$ in $t_1$ and again using the
Proposition \ref{Reliability function}, the proof follows.
\end{proof}

\begin{proof}[Proposition \ref{prop-int}]
The proof is immediate since
$
\textrm{E}[a^{-\perp}(X)]
=
{\beta\alpha^2}
\textrm{E}\left[X^2\right]/2
+
\beta
+
{\beta\alpha} \textrm{E}\big[X\sqrt{(\alpha X)^2+4}\big]/2
$
and since the function 
$x\mapsto x\sqrt{(\alpha x)^2+4}$
is odd.
\end{proof}

\begin{proof}[Theorem \ref{theorem-p}]
1. By \eqref{new-density} and Proposition \ref{Marginal} the conditional PDF of $T_1$ given $T_2$,
denoted by $f_{T_1|T_2}(t_1|t_2)$, is given by
\begin{align*}
f_{T_1|T_2}(t_1|t_2)
&=
\frac{1}{\sqrt{2\pi(1-\rho^2)}} \
\exp\left(-\frac{1}{2}\{c_{1,2}(t_1,t_2;\rho)\}^2\right) \, a_1'(t_1),
\end{align*}
where $c_{1,2}(t_1,t_2)$ is defined in \eqref{cp}. Hence,
\begin{align*}
\textrm{E}(T_1|T_2=t_2)
=
\int_{0}^{\infty} t_1 \,\frac{1}{\sqrt{2\pi(1-\rho^2)}}
\exp\left(-\frac{1}{2}\{c_{1,2}(t_1,t_2;\rho)\}^2\right) \, a_1'(t_1) \, \textrm{d}t_1
=
\textrm{E}\left[a_1^{-\perp}\left(\sqrt{1-\rho^2} \, Z+\rho a_2(t_2)\right)\right],
\end{align*}
where $Z\sim \textrm{N}(0,1)$.
Since $\sqrt{1-\rho^2}\, Z+\rho a_2(t_2)\sim N\big(\rho a_2(t_2), \, 1-\rho^2\big)$,
by Proposition \ref{prop-int}, the right-hand side of the above equality is
\[
\frac{\beta_1\alpha_1^2}{2}
\left[
\frac{2}{\alpha_1^2}+ (1-\rho^2)+ \rho^2 a_2^2(t_2)
\right]
\]
and the proof follows.

2.
By Proposition \ref{Marginal}, $T_k\sim\textrm{RBS}(\mu_k,\delta_k)$ for $k=1,2$.
Using properties of conditional expectation  and Item 1, we have
\begin{align}\label{ini-ig}
\textrm{E}[T_1T_2]
=
\textrm{E}\big[T_2\textrm{E}[T_1|T_2]\big]
=
\frac{\beta_1\alpha_1^2}{2}
\left[
\left(\frac{2}{\alpha_1^2}+ (1-\rho^2)\right)\textrm{E}[T_2]
+
\rho^2 \textrm{E}\big[T_2 a_2^2(T_2)\big]
\right].
\end{align}
By definition of $a_2(\cdot)$ and by Item 3 of
Subsection~\ref{pdf:RBS}, note that
\[
\textrm{E}[T_2]=\frac{\beta_2}{2}(\alpha_2^2+2),
\quad
\textrm{E}\big[T_2 a_2^2(T_2)\big]
=
\frac{1}{\alpha_2^2}\left(\frac{1}{\beta_2^2}\textrm{E}[T_2^2]+\beta_2-2\textrm{E}[T_2]\right)
=
\frac{\beta_2}{2}(2+3\alpha_2^2).
\]
Combining the equalities above with \eqref{ini-ig}, the proof follows.
Finally, since $T_k\sim\textrm{RBS}(\mu_k,\delta_k)$ for $k=1,2$, by Item 3 of
Subsection~\ref{pdf:RBS} and by Item 2 (Item 3),
the proof of Item 3 (Item 4) follows.
\end{proof}

\begin{proof}[Proposition \ref{Equilibrium distribution}]
The result immediately follows from \eqref{sf}, \eqref{equilibrium} and from
Theorem \ref{theorem-p} (Item 2).
\end{proof}

\begin{proof}[Proposition \ref{prop:3.9}]
The proof follows from Proposition \ref{prop:pre}.
\end{proof}

\begin{proof}[Theorem \ref{theo:3.3}]
The proof is direct by using the PDF in \eqref{sec2:02} and making suitable transformations.
\end{proof}



\begin{proof}[Theorem \ref{prop-failures}]
By Proposition 3.4 in \cite{Shaked} the \textrm{CHR}
$t_2\mapsto h(t_1|T_2>t_2)$ is decreasing if and only if the function
\begin{align*}
K_h(t_1,t_2)
&=
\int_{t_2}^{\infty}\int_{t_1}^{\infty}f_{T_1,T_2}(u,v)\, \textrm{d}u \, \textrm{d}v
\end{align*}
is $\textrm{TP}_2$,  and the \textrm{MRF}
$t_2\mapsto m(t_1|T_2=t_2)$ is increasing
if and only if the function
\[
K_m(t_1,t_2)=
\int_{t_1}^{\infty}\int_{v}^{\infty}f_{T_1,T_2}(u,t_2)\, \textrm{d}u \, \textrm{d}v
\]
is $\textrm{TP}_2$.
Then, by Theorem \ref{Holland and Wang} it is enough to verify that
$\gamma_{K_h}(t_1,t_2)$ and  $\gamma_{K_m}(t_1,t_2)$ are non-negative
functions. In what follows we shall prove this fact. By Proposition \ref{prop-pre} we rewrite the functions $K_h, K_m$ as
$$
K_h(t_1,t_2)
=
\int_{t_2}^{\infty} f(v;\mu_2,\delta_2)\big\{1-\Phi\big(c_{1,2}(t_1,v;\rho)\big)\big\} \, \textrm{d}v,
\quad
K_m(t_1,t_2)
=
f(t_2;\mu_2,\delta_2) \,
\int_{t_1}^{\infty}1-\Phi\big(c_{1,2}(v,t_2;\rho)\big) \,\textrm{d}v,
$$
where $c_{1,2}(v,t_2;\rho)$ is defined in \eqref{cp}. A straightforward calculus shows that
\begin{eqnarray*}
\gamma_{K_h}(t_1,t_2)
&=&
m(t_1,t_2;\rho)
\Bigg[
\left\{1- \Phi\big(c_{1,2}(t_1,t_2;\rho)\big)\right\}
\int_{t_2}^{\infty} f(v;\mu_2,\delta_2) \phi\big(c_{1,2}(t_1,v;\rho)\big) \,\textrm{d}v
\\
&&-
\phi\big(c_{1,2}(t_1,t_2;\rho)\big) \int_{t_2}^{\infty}
f(v;\mu_2,\delta_2)
\left\{
1
-
\Phi\big(c_{1,2}(t_1,v;\rho)\big)
\right\} \,\textrm{d}v
\Bigg],\\
\gamma_{K_m}(t_1,t_2)
&=&
\ell(t_1,t_2;\rho)
\Bigg[
\left\{1- \Phi\big(c_{1,2}(t_1,t_2;\rho)\big)\right\}
\int_{t_1}^{\infty} \phi\big(c_{1,2}(v,t_2;\rho)\big) \,\textrm{d}v
\\
&&-
\phi\big(c_{1,2}(t_1,t_2;\rho)\big) \int_{t_1}^{\infty}1
-
\Phi\big(c_{1,2}(v,t_2;\rho)\big) \,\textrm{d}v
\Bigg],
\end{eqnarray*}
where
\[
\begin{array}{c}
m(t_1,t_2;\rho)
=
\frac{1}{\sqrt{1-\rho^2}}
\frac{a_1'(t_1)f(t_2;\mu_2,\delta_2)}{[K_h(t_1,t_2)]^2},
\quad
\ell(t_1,t_2;\rho)
=
\frac{\rho}{\sqrt{1-\rho^2}}
\,
\frac{a_2'(t_2)}
{
\left[\int_{t_1}^{\infty}1
-
\Phi\big(c_{1,2}(v,t_2;\rho)\big) \,\textrm{d}v\right]^2
}.
\end{array}
\]
Since $a_k'(t_k)>0$, $m(t_1,t_2;\rho)>0$ and since $\rho>0$ we have $\ell(t_1,t_2;\rho)>0$.
Then, to prove that $\gamma_{K_h}(t_1,t_2)\geqslant 0$ and
$\gamma_{K_m}(t_1,t_2)\geqslant 0$, we must prove that
\begin{eqnarray*}
\frac{\int_{t_2}^{\infty} f(v;\mu_2,\delta_2) \phi\big(c_{1,2}(t_1,v;\rho)\big) \,\textrm{d}v}
{\phi\big(c_{1,2}(t_1,t_2;\rho)\big)}
&\leqslant&
\frac{\int_{t_2}^{\infty} f(v;\mu_2,\delta_2) \left\{1-\Phi\big(c_{1,2}(v,t_2;\rho)\big) \right\} \,\textrm{d}v}
{\left\{1- \Phi\big(c_{1,2}(t_1,t_2;\rho)\big)\right\}},\\
\frac{\int_{t_1}^{\infty} \phi\big(c_{1,2}(v,t_2;\rho)\big) \,\textrm{d}v}
{\phi\big(c_{1,2}(t_1,t_2;\rho)\big)}
&\geqslant&
\frac{\int_{t_1}^{\infty}1-\Phi\big(c_{1,2}(v,t_2;\rho)\big) \,\textrm{d}v}
{\left\{1- \Phi\big(c_{1,2}(t_1,t_2;\rho)\big)\right\}}.
\end{eqnarray*}
Let $Z\sim \textrm{N}(0,1)$.
To verify the inequalities above it is sufficient to prove that
\begin{eqnarray}
\frac{\textrm{P}\big(Z>c_{1,2}(t_1,v;\rho)\big)}
{\textrm{P}\big(Z>c_{1,2}(t_1,t_2;\rho)\big)}
&\geqslant&
\frac{\phi\big(c_{1,2}(t_1,v;\rho)\big)}{\phi\big(c_{1,2}(t_1,t_2;\rho)\big)},
\quad \textrm{for each} \ v>t_2,\label{inq-a}
\\
\frac{\textrm{P}\big(Z>c_{1,2}(v,t_2;\rho)\big)}
{\textrm{P}\big(Z>c_{1,2}(t_1,t_2;\rho)\big)}
&\leqslant&
\frac{\phi\big(c_{1,2}(v,t_2;\rho)\big)}{\phi\big(c_{1,2}(t_1,t_2;\rho)\big)},
\quad \textrm{for each} \ v>t_1.  \label{inq-b}
\end{eqnarray}
To verify the  inequality \eqref{inq-a} we will use the Gaussian tail inequality
\eqref{The Gaussian Tail Inequality}. In fact, since
$c_{1,2}(t_1,v;\rho)\geqslant c_{1,2}(t_1,t_2;\rho)\geqslant 1$
for all $v>t_2$ (by Items 1 and 2 of Lemma \ref{pre-lemma}),
using the inequality \eqref{The Gaussian Tail Inequality} we have
\begin{eqnarray*}
\frac{\textrm{P}\big(Z>c_{1,2}(t_1,v;\rho)\big)}
{\textrm{P}\big(Z>c_{1,2}(t_1,t_2;\rho)\big)}
&\geqslant&
\frac{ \left[\frac{1}{c_{1,2}(t_1,v;\rho)}-\frac{1}{c^3_{1,2}(t_1,v;\rho)}\right] \phi\big(c_{1,2}(t_1,v;\rho)\big)}
{\textrm{P}\big(Z>c_{1,2}(t_1,t_2;\rho)\big)}
\\
&\geqslant&
\frac{\left[\frac{1}{c_{1,2}(t_1,v;\rho)}-\frac{1}{c^3_{1,2}(t_1,v;\rho)}\right]
c_{1,2}(t_1,t_2;\rho)\,
\phi\big(c_{1,2}(t_1,v;\rho)\big)}
{\phi\big(c_{1,2}(t_1,t_2;\rho)\big)}
\\
&\geqslant&
\frac{\phi\big(c_{1,2}(t_1,v;\rho)\big)}{\phi\big(c_{1,2}(t_1,t_2;\rho)\big)},
\quad \textrm{for each} \, v>t_2,
\end{eqnarray*}
where in the third inequality we use the Item 4 of Lemma \ref{pre-lemma}.
Analogously, using the Gaussian tail inequality, Lemma \ref{pre-lemma}
(specifically, Items 1,3 and 5), the proof of inequality \eqref{inq-b} follows. To prove the Item 3 it is enough to use the identity
$\int_{0}^{t_1}h(t|T_2>t_2)dt=-\log S(t_1|T_2>t_2)$ and the Item 1.
Finally, by \eqref{tp2} and Theorem \ref{Holland and Wang},
the \textrm{PDF} $f_{T_1,T_2}(t_1,t_2)$
is $\textrm{TP}_2$ $(\textrm{RR}_2)$ when $\rho>0$ $(<0)$, then, by \cite{Shaked} the
$\textrm{CHR}$ of $(T_1|T_2=t_2)$ is decreasing (increasing), completing the proof of Item 4.
\end{proof}
\begin{proof}[Theorem \ref{theo:4.1}]
Let ${\bm T}=(T_{1},T_{2})^{\top}$ follow
a
$T\sim\textrm{BRBS}({\bm \theta})$,
then
\begin{eqnarray*}
&\textrm{Var}[T_{k}]=\frac{\mu_{k}^{2}(2\delta_{k}+5)}{(\delta_{k}+1)^{2}},
\quad
\textrm{Var}[T_{k}^{-1}]=\frac{(2\delta_{k}+5)(\delta_{k}+1)^2}{\mu_{k}^{2}\delta_{k}^{4}},
\quad
\textrm{Cov}[T_{k}]=1-\frac{(\delta_{k}+1)^2}{\delta_{k}^{2}},
\quad k=1,2.
\end{eqnarray*}
Consider $S_{k}=\sum_{i=1}^{n}T_{kj}/n$ and $R_{k}^{*}=R_{k}^{-1}=\sum_{i=1}^{n}({1}/{T_{ki}})$,
with $k=1,2$, which implies that the vector $(S_{k},R_{k}^{-1})^{\top}$ is
bivariate normal distributed, that is,
\begin{eqnarray*}
\sqrt{n}
\left(
\begin{array}{c}
S_{k}-\textrm{E}[T_{k}]\\
R_{k}^*-\textrm{E}[T_{k}^{-1}]
\end{array}
\right)
\sim
\textrm{N}
\left[
\left(
\begin{array}{c}
0\\
0
\end{array}
\right),
\left(
\begin{array}{cc}
\textrm{Var}[T_{k}], &1-\textrm{E}[T_{k}]\textrm{E}[T_{k}^{-1}]\\
1-\textrm{E}[T_{k}]\textrm{E}[T_{k}^{-1}], &\textrm{Var}[T_{k}]
\end{array}
\right)
\right].
\end{eqnarray*}
However, we need to find the asymptotic joint distribution of $(\widetilde\mu_{k},\widetilde\delta_{k})^{\top}$. Consider
$\widetilde\mu_{k}=f_{k}(S_{k},R_{k}^{*})$ and $\widetilde\delta_{k}=f_{k}(S_{k},R_{k}^{*})$, such that
$f_{k}(x,y)=x$ and $f_{2}(x,y)=1/(\sqrt{x{y}}-1)$.
By using the Delta method, we readily have
\begin{eqnarray*}
\sqrt{n}\left(
\begin{array}{c}
\widetilde\mu_{k}-\mu_{k}\\
\widetilde\delta_{k}-\delta_{k}
\end{array}
\right)
\sim
\textrm{N}
\left(
\left[
\begin{array}{c}
0\\
0
\end{array}
\right],
{\bm\Sigma}_{k}
\right),
\quad
\textrm{where}
\quad
{\bm\Sigma}_{k}
=
\left(
\begin{array}{cc}
\frac{\mu_{k}^{2}(2\delta_{k}+5)}{(\delta_{k}+1)^2} & -\frac{2\mu_{k}\delta_{k}}{\delta_{k}+1}\\
-\frac{2\mu_{k}\delta_{k}}{\delta_{k}+1} & 2\delta_{k}^{2}\\
\end{array}
\right),
\quad k=1,2.
\end{eqnarray*}
\end{proof}


\end{document}